\documentclass[manuscript]{acmart}

\AtBeginDocument{%
  \providecommand\BibTeX{{%
    \normalfont B\kern-0.5em{\scshape i\kern-0.25em b}\kern-0.8em\TeX}}}

\setcopyright{acmlicensed}
\copyrightyear{2024}
\acmYear{2024}
\acmDOI{XXXXXXX.XXXXXXX}

\acmJournal{JATS}
\acmVolume{37}
\acmNumber{4}
\acmArticle{111}
\acmMonth{8}

\acmISBN{978-1-4503-XXXX-X/18/06}

\usepackage{multirow}
\usepackage{algorithm}
\usepackage{algpseudocode}

\usepackage{amsthm}
\usepackage{mathtools}          
\usepackage{mathrsfs}           
\usepackage{graphicx}           
\usepackage{subcaption}         
\usepackage[space]{grffile}     
\usepackage{url}                
\usepackage{caption}
\usepackage{soul}               
\usepackage{paralist}
 \usepackage{multirow}

\usepackage{enumitem}
\setlist[itemize]{noitemsep, topsep=0pt}

\newcommand{\Fc}{\mathcal{F}}

\expandafter\newcommand\csname r@tocindent4\endcsname{4in}
\begin{document}

\title{Cooperation and Fairness in Multi-Agent Reinforcement Learning}

\author{Jasmine Jerry Aloor}
\email{jjaloor@mit.edu}
\orcid{0000-0003-4257-3870}
\author{Siddharth Nayak}\orcid{0000-0003-4663-8045}
\email{sidnayak@mit.edu}
\author{Sydney Dolan}
\email{sdolan@mit.edu}
\orcid{0009-0002-8620-0005}
\author{Hamsa Balakrishnan}
\email{hamsa@mit.edu}
\orcid{0000-0002-8624-7041}
\affiliation{%
  \institution{Massachusetts Institute of Technology}
  \streetaddress{77 Massachusetts Avenue}
  \city{Cambridge}
  \state{Massachusetts}
  \country{USA}
  \postcode{02139}
}

\renewcommand{\shortauthors}{Aloor, et al.}

\begin{abstract}

Multi-agent systems are trained to maximize shared cost objectives, which typically reflect system-level efficiency. However, in the resource-constrained environments of mobility and transportation systems, efficiency may be achieved at the expense of fairness --- certain agents may incur significantly greater costs or lower rewards compared to others. Tasks could be distributed inequitably, leading to some agents receiving an unfair advantage while others incur disproportionately high costs. It is, therefore, important to consider the tradeoffs between efficiency and fairness in such settings. 

We consider the problem of fair multi-agent navigation for a group of decentralized agents using multi-agent reinforcement learning (MARL).  
We consider the reciprocal of the coefficient of variation of the distances traveled by different agents as a measure of fairness and investigate whether agents can learn to be fair without significantly sacrificing efficiency (i.e., increasing the total distance traveled). We find that by training agents using min-max fair distance goal assignments along with a reward term that incentivizes fairness as they move towards their goals, the agents (1) learn a fair assignment of goals and (2) achieve almost perfect goal coverage in navigation scenarios using only local observations. For goal coverage scenarios, we find that, on average, the proposed model yields a 14\% improvement in efficiency and a 5\% improvement in fairness over a baseline model that is trained using random assignments. Furthermore, an average of 21\% improvement in fairness can be achieved by the proposed model as compared to a model trained on optimally efficient assignments; this increase in fairness comes at the expense of only a 7\% decrease in efficiency. Finally, we extend our method to environments in which agents must complete coverage tasks in prescribed formations and show that it is possible to do so without tailoring the models to specific formation shapes.
\href{https://github.com/Jaroan/Fair-MARL}{[Code]}\footnote{ \href{https://github.com/Jaroan/Fair-MARL}{Code base: https://github.com/Jaroan/Fair-MARL}}

\end{abstract}




\maketitle

\section{Introduction}
\label{sec:introduction}
Multi-agent vehicular systems, where large numbers of vehicles coordinate to execute complex missions, have the potential to transform the transportation and mobility domains. Such systems have wide-ranging applications \cite{Schranz-swarm-survey-2020}, including disaster response \cite{chin-JAXA-2023}, wildfire detection \cite{Allison-wildfire}, sensing and monitoring \cite{mersheeva-surveillance,suduwella-mosquitoes,Balasingam-DroNet-2021}, ridesharing \cite{vrp-maxtpt}, agriculture \cite{farmbeats}, and spacecraft operations \cite{bandyopadhyay-formation-spacecraft-2016,dolan2023satellite}. Despite the differences in application domains, these operations tend to occur in resource-constrained environments where it is important to make efficient use of resources; however, the quest for efficiency alone in these situations can often mean that fairness is sacrificed. In other words, some agents (vehicles or users) may receive significantly better or worse outcomes relative to others \cite{Balasingam-Mobisys-2021,Chin_2023}. We require methods that can guide agents to achieve the desired behavior of efficiency and fairness.

Multi-agent reinforcement learning (MARL) approaches enable the investigation of a range of interactions (e.g., competitive or cooperative) between agents or teams of agents. The ability of reinforcement learning (RL) to learn by trial and error makes it well-suited for problems in which optimization-based methods are not effective. In particular, MARL approaches are better suited to these problems due to their fast run times, superior performance, and ability to model cooperative behavior through shared goals between agents using appropriate reward structures in complex, unstructured environments. One common approach to training cooperative multi-agent reinforcement learning models is the centralized training decentralized execution (CTDE) paradigm. In CTDE, decentralized agent policies are trained through a centralized mixing model with global state information, while agents select actions using only their local observations \citep{zhou2023centralized}. As an example, agents could be required to complete a set of tasks in a collaborative manner, where agents share key information with others that help the system accomplish the tasks optimally. Each agent could be assigned a particular task using a centralized method. The learning process occurs as the agents seek to find an optimal policy to maximize the shared reward function. While this shared reward encourages collaboration amongst agents, it does not consider the fairness of each agent's task assignment, meaning that certain agents could contribute towards a disproportionate portion of the overall shared reward. We wish to avoid any assignment that distributes the tasks in an inequitable way that leads to some agents receiving an unfair advantage while others starve for resources.

In this paper, we focus on multi-agent navigation and collision avoidance problems in which there are $N$ agents trying to cooperate with each other to collectively solve a task in a 2D environment with static and dynamic obstacles. 
The overarching objective is for all the agents to complete their tasks in the shortest time possible and avoid collisions with other agents and obstacles while also maintaining fairness across all agents in the team. Through cooperative decentralized multi-agent reinforcement learning, we study the problem of learning \emph{fair}, distributed policies. This problem setting is quite general and arises in many contexts, e.g., search-and-rescue robot teams \cite{search_rescue}, environmental monitoring \cite{env_monitoring}, formation control \cite{formation_1, graph_formation_marl}, and drone delivery systems \cite{drone_delivery1, drone_delivery2, drone_delivery3}. We also examine the extensibility of our algorithm to formation-based tasks. Convoy formation has been studied in the context of a number of autonomous transportation modes, including railways \cite{Henke2006,Henke2008}, roadways \cite{Heinovski2018, Johansson2022,AdlerMK-wafr16} and advanced air mobility \cite{Ishihara-2021}, because of the potential savings in terms of operator workload \cite{McKinsey2018} and fuel reduction \cite{Tsugawa2016}. For this reason, we extend our methodology to settings in which the agents may need to coordinate into a prescribed formation. 

The overarching objective of this work is the development of a method that will allow multiple agents in an environment to navigate and complete coverage tasks in a scalable, efficient, and fair manner. Our solution approach focuses on the evolution of fairness throughout the course of an episode in a cooperative, decentralized multi-agent reinforcement learning setting. Specifically, we design a reward function in training that encourages both the completion of the task and the overall fairness in each agent's task selection and execution. The proposed reward is based on the combination of a min-max fair distance goal assignment along with a term that incentivizes the agents to maintain fairness while moving to their goals. 
 The main contributions of this paper are:
\begin{itemize}
    \item We introduce a reward function that enables the underlying MARL algorithm to learn fair behavior as agents navigate to complete coverage tasks.
    \item We demonstrate a decentralized learning-based goal assignment approach that allows agents to adaptively select their goals during execution.
    \item We show that our method scales to allow an arbitrary number of agents to create any formation shape without the need for retraining. 
\end{itemize}

The remainder of this paper is organized as follows: Section \ref{sec:related} reviews related work. Section \ref{sec:methods} introduces the problem formulation and methods, while Section \ref{sec:results} presents and discusses the results of evaluating our method across a range of scenarios, both random and formation-based. We conclude the paper in Section \ref{sec:conclusions} with promising directions for further investigation. 

\section{Related Work}

\label{sec:related}

\subsection{Fairness}

Fairness has been extensively studied in many contexts, including in game theory \citep{game_theory_fair_survey}, economics \citep{economic_fairness}, and machine learning \citep{ML_fairness}. In optimization approaches, fairness is usually formulated as a strict mathematical objective or constraint. These approaches work better when the problem structure is fixed, such as a predefined network structure or creating specific formations \cite{tzikas2024distributed}. In machine learning, work in fairness often refers to mitigating social biases and the social, legal, and ethical aspects of machine learning discrimination.  One of the most common classes of approaches is the alpha fairness method \citep{alpha_fairness}. Our work is concerned with fairness in the network engineering sense \citep{network_fairness_def}, where individual users receive a fair share of system resources  \citep{Kleinberg_Rabani_Tardos_2001}.  

There are relatively few works in MARL studying this definition of fairness. It is a growing area of interest due to its impact on shaping cooperative behavior in resource-constrained environments.  \citet{de2008priority} incorporated priority awareness into their fairness modeling. As human notions of fairness often consider it fair if agents receive slightly different rewards in the presence of additional contextual information, this work assumes that each agent knows the true priority of all other agents in the simulation. This approach produces sub-optimal reward solutions. Unlike the priority-based model, \citet{jiang2019learning} propose a Fair-Efficient Network (FEN) algorithm to maintain fairness and efficiency using a hierarchical controller that chooses between efficient and fair policies based on the average utility obtained at every episode. The authors state that learning efficiency and fairness is challenging with a single policy. We overcome this using our optimization-based min-max fair assignment, which finds a fair assignment using the distance metric indirectly considering efficiency as well.

The Fair-Efficiency Multi-Agent Deep Deterministic Policy Gradient (FE-MADDPG) algorithm \cite{liu2022fairness} implements a fairness reward similar to FEN. However, it operates using global information, offering lower levels of privacy and decentralization. Another key limitation is scaling this method to different numbers of entities, which would require retraining the model every time.
\citet{zimmer2021learning} propose a fairness framework that balances efficiency and equity using a welfare function adaptable to various fairness metrics (lexicographic maximin, Gini). Their approach includes sub-networks to account for fairness and efficiency, where the agents choose a self-oriented efficient policy or a fair team-oriented policy based on a probability parameter that decreases over time. Initially, the agents are trained to maximize efficiency, and later, they are trained to improve fairness. Our approach trains agents to navigate to goals based on a min-max fair assignment, aiming to equalize costs associated with navigating to goals without sacrificing efficiency.

\citet{grupen2022cooperative} find that sophisticated coordination behavior only emerges when there is a shared reward, but this emergent behavior does not ensure fairness. They introduce a soft-constraint equivariant policy learning method to dynamically balance the fairness-utility trade-off.  In contrast to \citet{de2008priority}, \citet{liu2022fairness} and \citet{grupen2022cooperative}, our work considers fairness in navigation-based settings where global information is not available to each agent.

\subsection{Communication Structures in Multi-Agent Reinforcement Learning}
Our work focuses on the problem of multi-agent navigation and collision avoidance among a set of decentralized $N$ agents that can only sense the presence of other obstacles and agents within a limited radius $r$. Communication between agents is vital for them to complete their respective tasks successfully. The study of the role of communication between agents is an active and extensive field within MARL, so we refer the reader to a survey on the topic \citep{Zhu_Dastani_Wang_2024} for a comprehensive description. We highlight several works in this area that address problems similar to ours.

Existing MARL work on this problem often assumes that even if the behavior of the agents is decentralized, communication amongst them is \textit{centralized}. This means that all agents have access to messages from all other agents in the environment. Unfortunately, this means that as the number of agents increases, the computational expense of communication increases superlinearly. Centralized communication also creates privacy concerns, where agents are unable to participate in the MARL framework without sharing data with everyone. To address this expense, ATOC \citep{ATOC, TarMAC} relies on attention mechanisms to provide a compact representation of message priority. In EMP \citep{EMP}, the environment is translated into a shared agent-entity graph representation that allows agents to communicate along connected edges.  This formulation provides a compact graph representation of the communication connections between all entities in the environment. However, all agents must know the positions of all entities in the graph at the beginning of an episode, and thus, a similar centralized communication constraint arises. In contrast to these centralized communication approaches, InforMARL \citep{informarl} differs from these works in that it relies on only locally available information throughout training and during evaluation. Our algorithm relies on InforMARL as the underlying MARL algorithm for training and evaluation, with several adaptations to its reward and buffer structure. 

\subsection{Formation in MARL}
Multi-agent reinforcement learning has proven effective for the formation control of multiple robots. Shen et al. \cite{digital_twin_flocking} propose a modified deep deterministic policy gradient (DDPG) algorithm to achieve the formation control for multiple unmanned aerial vehicles with the help of digital twin technology. Graph convolutional networks are used for policy gradient methods in \cite{gpg}, which leverage the underlying graph formed by the agents. To maintain the formation shape while navigating obstacles in the environment, \citet{formation_1} use a hybrid approach where they switch between an RL controller and a PID controller based on the state. However, the above-mentioned works use some form of reward shaping where they only account for goal-reaching and collision avoidance as their main goals. These works assume the formation pattern to be fixed and task each agent to navigate to a fixed position in the environment. Our work contrasts these methods by introducing a shape-independent approach that allows agents to come into various formations in a scalable manner. We do not fix the assignment of agents to positions of the formations; instead, agents decide what position to occupy based on local observations and updated rewards at each time step.

\section{Methodology}
\label{sec:methods}

In this section, we outline the methodologies used to achieve policies with fairer outcomes for agents navigating to goals. We describe the environment and the training setup, detail the formulations employed for agent observations, and explain the modifications made to the reward functions that lead to fair behavior. We also discuss centralized efficiency and fairness optimization formulations for goal assignments.

We train an adapted version of InforMARL, an existing CTDE multi-agent reinforcement learning algorithm, on our modified reward functions. The details of this algorithm are beyond the scope of this paper and can be found in \cite{informarl}.

\subsection{Preliminaries}
Following the InforMARL framework, our environment comprises entities categorized into agents, obstacles, and goals. For each agent $i$ at each time-step $t$,  we define an agent-entity graph as $g^{(i)}_t \in \mathcal{G}: (\mathcal{V}, \mathcal{E})$, where each node $v\in\mathcal{V}$ is an entity in the environment. Entities are connected to each other by edges if they are within a certain sensing distance. Agent-agent edges are bi-directional, which is equivalent to a communication channel between them, whereas agent-non-agent edges are unidirectional, with messages being passed from the non-agent entity to the agent.

Our environment is based on the Multi-Particle Environments (MPE) \citep{MAPE} collection of tasks. We formulate our environment as a Decentralized Partially Observable Markov Decision Process (Dec-POMDP) defined by the tuple $\langle N, S, O, \mathcal{A}, \mathcal{G}, P, R,\gamma \rangle$, where:
\begin{itemize}
    \item $N$ is the number of agents
    \item $s\in S = \mathbb{R}^{N\times D}$ is the state space of the environment, with $D$ as the dimension of the state
    \item $o^{(i)}=O(s^{(i)})\in \mathbb{R}^{d}$ is the local observation for agent $i$ 
    \item $a^{(i)} \in \mathcal{A}$ is the action space for agent $i$. The action space for each agent is discretized such that it can control unit acceleration and deceleration in the $x$- and $y$- directions
    \item $g^{(i)} \in \mathcal{G}(s; i)$ is the graph network formed by the entities in the environment with respect to agent $i$
    \item $P(s'|s, A)$ is the transition probability from $s$ to $s'$ given the joint action $A$
    \item $R(s, A)$ is the joint reward function
    \item $\gamma \in [0,1)$ is the discount factor
\end{itemize}
We investigate a coverage navigation problem \citep{targetCoverage1, targetCoverage2} where a group of agents must navigate to different goals. We have $N$ agents that navigate a 2D space using a double integrator dynamics model \citep{double_integrator_model}.  There are $N$ goals in the environment, each denoted by $\zeta^{(i)}$. Agents navigate in such a way that each agent reaches a unique goal while avoiding obstacles and walls. The task is to find a policy $\Pi= \left(\pi^{(1)}, \cdots, \pi^{(N)}\right)$, where $\pi_{\theta}^{(i)}\left(a^{(i)}|o^{(i)}, g^{(i)}\right)$ for agent $i$ selects action $a^{(i)}$ based on its graph network $g^{(i)}$ and the local observation $o^{(i)}$.

\subsection{Agent training framework}
\label{ssec:agent_training_framework}

We group our training steps into multiple episodes. At the beginning of each episode, the environment is initialized with agents starting at random locations and goals randomly distributed in the environment. The policy provides an action for each agent $i$ in the first step. The agent executes the action and observes the change in state. At each time step, a goal assignment mechanism assigns a goal to the agent, and a preliminary distance-based reward $\mathcal{R}_\mathrm{dist}(s_t, a^{(i)}_t)$ is computed by taking the negative of the Euclidean distance to this assigned goal. In Section \ref{ssec:goal_assignment}, we discuss the goal assignment techniques used in this work. Additionally, we include a fairness metric to the reward function to create a "fair" reward. In Section \ref{ssec:fairness_section}, we discuss the structure of our fairness reward function.

The agent uses its sensing radius to detect the presence of other entities in its vicinity. A graph $g^{(i)}_t$ is created using these entities and passed into a graph neural network (GNN). The output from the GNN--$x^{(i)}_{agg}$--and the observation vector $o^{(i)}$ are used by the agent to select its next action. In Section \ref{ssec:observations}, we elaborate on the observation functions and the input to the graph network. Agents are penalized when they collide with other agents or obstacles, with a collision penalty of $-C$ added to the reward. Agents utilize the observations and rewards to learn which goal locations to navigate to. When an agent reaches its assigned goal, it receives a one-time goal reward $\mathcal{R}_\mathrm{goal}(s_t, a^i_t)$. These agents are flagged as `done' and remain stationary at the goal until the end of the current episode. We also restrict the actions that can be taken by these `done' agents. This prevents the agents from drifting away under the collision force influence of other agents that are still navigating to their respective goals. We call this setting `death-masking' based on \cite{mappo}. When all agents arrive at their goals and are `done', we end the episode and restart the process.

\begin{figure*}[t]
    \centering
    \includegraphics[width=\textwidth]{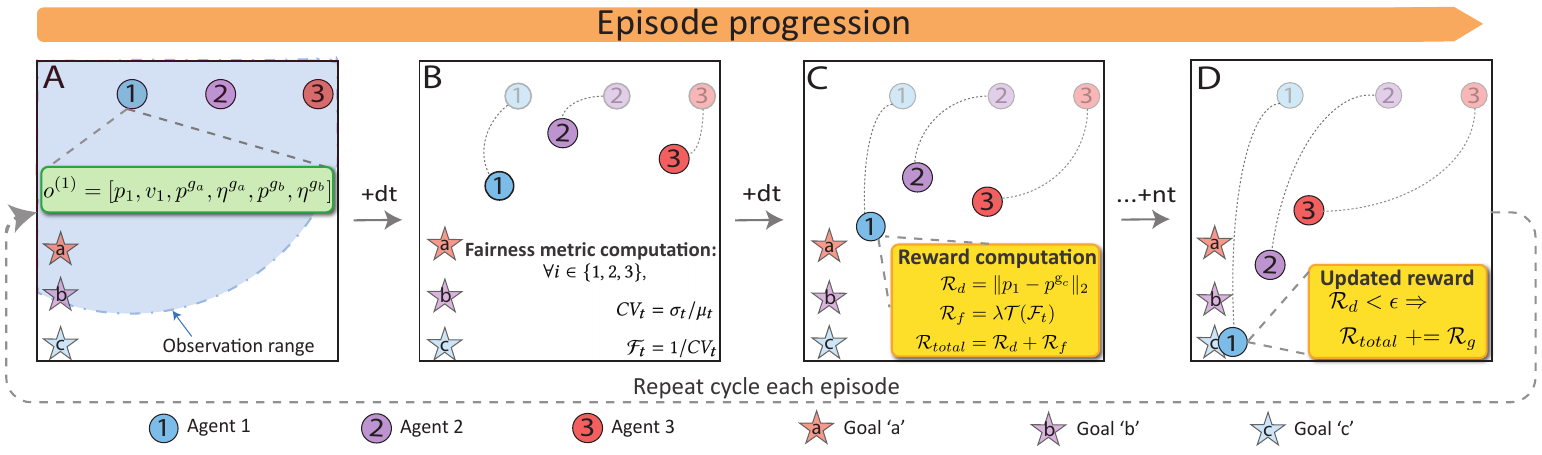}
    \caption{Overview of the training: In the navigation scenario, we track the path of the agents as an episode progresses. Frame A: The episode starts with agents and goals initialized to random positions. For ease of representation, we have ordered them along two lines. The observation vector of agent 1, $o^{(1)}$ is shown in the green box along with its observation radius highlighted in the blue shaded circle. Frames B and C: At every time step, for each agent, the fairness metric $\mathcal{F}_t$ is computed along with each agent's rewards. The agents are assigned goals based on an optimal or fair distance cost. Frame D: Once an agent reaches the assigned goal, it is given a goal reward $\mathcal{R}_\mathrm{goal}$ and is flagged "done" for that episode.}
    \label{fig:training-overview}
    \Description[Overview of the training]{Overview of the episode progression during training with four frames showing various steps involved}
\end{figure*}

Figure \ref{fig:training-overview} shows an example of the sequence of steps followed within a training episode. In Frame A, the agents are initialized on the top right of the environment, and the goals are placed on the bottom left. As the episode progresses, agents take action and observe the environment and changed states. The assigned goal information is provided to the agent based on the value of the distance-based reward $\mathcal{R}_\mathrm{dist}(s_t, a^{(i)}_t)$. At every time step, we can compute the fairness reward to be appended to the total reward for each agent, as shown in Frames B and C. When an agent reaches its goal, Frame D demonstrates that the agent receives an additional goal-reaching reward $\mathcal{R}_\mathrm{goal}(s_t, a^i_t)$.

\subsubsection{Fairness metric}
\label{ssec:fairness_section}
Fairness is context-dependent, and there are different definitions of fairness for different problems.
In this work, we focus on fairness in the context of multi-agent navigation to goal assignments. 
Prior works have considered fairness in multi-agent networked systems \cite{network_fairness_def, Balasingam-Mobisys-2021}. We extend the definition of fairness used in networked systems and choose the distance traveled and time each agent takes to reach its goal as the resources to be treated fairly. We want to minimize the standard deviation of distance traveled and time to reach goals among all agents. By reducing the spread of these metrics and incorporating these metrics in our reward function, we can add fairness to agent behavior in the system.
Large imbalances in travel distances could lead to significant disparities in the effort required by different agents. Our metric aligns with this task by ensuring a balanced workload across all agents, minimizing the variance in their travel distances.

The distance traveled by each agent $i$ at each time step $t$ is represented by $d^{(i)}_t$, with an overall mean $\mu_t$ and standard deviation $\sigma_t$ for all agents. We compute the coefficient of variation as 
${CV}_t = {\sigma_t}/{\mu_t}$. We choose our fairness metric $\mathcal{F} = 1 /CV$ to be the inverse of $CV$ so that it is a non-dimensional quantity with higher numerical values indicating greater fairness. To avoid a zero denominator, we include a small positive number $\epsilon$ to the fairness metric $\mathcal{F}$. Thus, the fairness metric is
\begin{equation} \label{eqn:fairness}
\begin{split}
   \mathcal{F}_t(s_t, a^{(i)}_t) &= \frac{\mu_t}{\sigma_t + \epsilon}
\end{split}
\end{equation}
We then pass $\Fc$ to every agent policy $\pi^{(i)}$ through a fairness reward function (\ref{eq:fair_reward}). The fairness metric is transformed through a hyperbolic tangent function whose zero value is shifted appropriately by $\tau_0$ to enable agents to have a higher fairness metric. This value is scaled up by a factor $\lambda$ to determine the fairness reward $\mathcal{R}_\mathrm{fair}(s_t, a^{(i)}_t)$. The overall reward is tuned with the addition of a fairness reward (FR).

 \begin{equation}
     \mathcal{R}_\mathrm{fair}(s_t, a^{(i)}_t) = \lambda \mathrm{tanh}(\mathcal{F}_t(s_t, a^{(i)}_t) - \tau_0)\label{eq:fair_reward}.
 \end{equation}

We investigated an alternate metric involving the standard deviation of fairness or its inverse, $1/\sigma_t$. Further analysis found that the trends in fairness are similar between the two metrics. Additionally, $1/\sigma_t$ alone does provide enough insight regarding the extent of fairness of a system when deployed over multiple scales. Thus, we continue our implementation with $1/CV$, which is a non-dimensional quantity and more extensible to a wider range of applications. The agents have a common objective to reach the targets, and the fairness metric does not directly conflict with the goal reward.

\subsubsection{Goal assignment}
\label{ssec:goal_assignment}
We assign agents to goals in training for the models to learn fair and efficient behaviors. Assigning agents to goals is a well-investigated problem \citep{goal_assign_survey}. When we trained models without goal assignments, the agents did not learn to cover unique goals.   Instead, we utilize random assignment, which provides increased fairness, an optimal distance-based cost assignment, which increases efficiency, and a fair assignment that uses min-max fairness that improves fairness while considering the distance cost.
In the optimal and fair assignment models, we calculate the assignment at every timestep. This ensures the agents maintain efficiency or fairness throughout the episode. 

The assignment function takes into account each agent's current position and the location of all goals in the environment. Note that these assignments are only used to determine the rewards for the agents and are not used to determine the goal positions for the agent's local or graph observation vectors.

\paragraph{Random assignment}
\label{sssec:random_assign}
Randomization is a commonly used resource allocation method to improve fairness. We assign agents to goals randomly at the start of the episode. We denote this assignment by (RA). At every time step, the agents are provided a reward ${\mathcal{R}}^{\mathrm{RA}}_\mathrm{dist}(s_t, a^{(i)}_t)$ based on the distance to these assigned goals. The random assignment does not consider any cost in the allocation and consequently may be very inefficient. In this paper, we use models trained with random goal assignments as the baseline. Randomness inherently offers a level of fairness by allowing the assignment of goals to agents without any bias and allows the exploration of diverse behaviors. It serves as a baseline to show how our proposed model outperforms a simple, fair-by-chance allocation strategy using the same setup.

\paragraph{Optimal distance cost assignment}
\label{sssec:opt_cost_assign}
For an efficient assignment, we use the linear sum assignment or minimum weight matching in the bipartite graphs algorithm. This matches each agent $i$ to a goal $j$ so that the total cost $c_{i j}$ for all agents is minimized, which here corresponds to minimizing the total distance, or $\min_{x_{ij}} \sum\limits_{i=1}^n \sum\limits_{j=1}^n c_{i j} x_{i j}$, where $x_{i j}$ is 1 if agent $i$ is assigned to goal $j$ and 0 otherwise. With the optimal distance cost assignment, we attempt to minimize the total distance traveled $D$ by each agent based on the locations of agents and goals. This distance-based cost approach does not consider fairness. We denote this assignment in our experiments by (OA). The distance-based reward for an optimally assigned goal is represented as ${\mathcal{R}}^{\mathrm{OA}}_\mathrm{dist}(s_t, a^{(i)}_t)$. Comparing against this model allows us to demonstrate the tradeoffs in efficiency while attempting to improve fairness.

\paragraph{Min-max fair assignment}
\label{sssec:minmax_assign}

We compare the optimal cost assignment with the performance of agents when the system is provided with a fair assignment of agent-goal results. Min-max fairness is a popular concept that reduces the worst-case cost in the assignment. Min-max fair assignment, often used in networking bandwidth allocation \cite{max_min_fair_sharenetworks}, and transportation scheduling problems \cite{GOLDEN1997445}, among others, prevents a single agent from being disproportionately overloaded.
We determine a min-max fair assignment by optimizing the objective $\min z$ subject to constraints, ensuring each agent and goal is assigned exactly once. Here, $z$ represents the maximum cost assigned to any agent, $c_{ij}x_{ij} \leq z$ where $c_{i j}$ and $x_{i j}$ are defined as above. At each iteration of the min-max fair assignment, the worst-off agent is given priority to get a better outcome.  With the min-max fair assignment, we attempt to minimize the maximum distance traveled by an agent based on the locations of agents and goals. We denote this type of assignment in our experiments by (FA). The distance-based reward for a fair assigned goal is represented as ${\mathcal{R}}^{\mathrm{FA}}_\mathrm{dist}(s_t, a^{(i)}_t)$.
This fair assignment differs from the fairness metric we use to evaluate our performance.

We explored the possibility of keeping a fixed assignment throughout the episode.
Optimization-based approaches to fairness in multi-agent systems assign goals \textit{a priori} to ensure a fair distribution of rewards amongst agents.
For team-based scenarios where individual agents must collaborate to solve a task collectively, having a fixed assignment-based approach to agent task selection can limit agents’ mobility. In the coverage navigation problem, a fixed assignment calculated at the start of the episode prevents them from capitalizing on opportunistic navigation changes that can lead to a desired behavior. Due to the non-stationarity of the environment, fixed assignment-based approaches can provide infeasible solutions, as they assume a centralized oracle directing agents to individual goals based on an initial environment setup. 
Finally, in realistic scenarios, it may be impossible to have a centralized oracle assigning goals for agents. In our approach, we rely on a decentralized system to solve this problem. We use local observations to inform agents of the presence of other agents and the availability of goals.

\subsubsection{Agent observations}
\label{ssec:observations}
Each agent's local observation vector consists of its position $p_i$ and velocity $v_i$ in a global frame of reference and the information about the locations of goals. Any goal position that is input to the agent's observation is the relative position of the goals with respect to the agent's position, i.e., $p^{\textrm{goal}_j}_i$.

In centralized MARL applications, agents are provided with all goal positions as well as information about the neighboring agents in the environments. This approach is not feasible for a scalable decentralized system, as it fixes the model's input and output matrices. This also leads to a lower level of privacy and assumes global knowledge of every agent's position.
When agents are sparsely distributed in larger environments, they can only observe parts of the environment, preventing them from knowing the positions of all goals \textit{a priori}. We model this in our observation function by only using the positions of the closest two goals from each agent.

We investigated the effects of not providing any goal information in the agent's observation, which led to agents wandering in the environment and not being able to reach goals. When only the closest goal's position was provided, agents would tend to loiter around that goal, even if it was occupied, without awareness of additional goals. However, increasing the amount of goal information too much increases the computational burden and decreases scalability. Moreover, it fixes the network input size, limiting flexibility when the number of goals changes dynamically. Balancing the amount of goal information provided allows agents to operate efficiently without compromising scalability or privacy.

We provide a goal occupancy flag in the observation vector that informs agents how close any agent is to that goal. It allows agents to know if a goal in their sensing range is occupied or will soon be occupied due to another agent's presence in its proximity. The goal occupancy flag $\eta$ is created for each goal in the environment. We populate the flags based on the distance of the closest agent to the goal. At the start of an episode, for each goal $\zeta^{(j)}$ with position ${p^{(j)}_g}$, we initialize all $\eta^{\textrm{goal}_j}$ to 0. As agents move closer to the goals, we calculate the minimum distance any agent is from a particular goal $j$,
\begin{equation}
\begin{aligned}
        d^{(j)}_{\textrm{min}} &= \min_{ i \in N } \lVert {x^{(i)} - {p^{(j)}_g}} \rVert_2 \\
    \eta^{\textrm{goal}_j}  &= 1 - d^{(j)}_{\textrm{min}}
\end{aligned}
\end{equation}
We restrict the value of $\eta$ to be within 0 and 1 for ease of computation. For a given goal, the flag value increases from 0 to 1 as an agent tries to reach it. In particular, $\eta^{\textrm{goal}_j}$ = 0 means the goal $\zeta^{(j)}$  is available for any agent, and 1 indicates that the goal is fully occupied. The goal occupancy flag update scheme is key to ensuring agents share accurate information. It also demonstrates agent cooperation.

When an agent is near all occupied goals and cannot sense the presence of other unoccupied goals in its vicinity, the agent slows down and explores the environment until it finds an available goal. We hasten this process by allowing the agent to request the position of the nearest unoccupied goal to prevent agents from slowing down near goals already occupied by other agents. Thus, the agent is explicitly made aware of the presence of goals outside its sensing range only when there are no nearby unoccupied goals.
The final ego observation vector $o^{(i)}$ can be represented as $o^{(i)} = [{p_i}, {v_i}, p^{\textrm{goal}_1}_i, \eta^{\textrm{goal}_1}_i, p^{\textrm{goal}_2}_i, \eta^{\textrm{goal}_2}_i ]$. 

Within an agent's sensing range, there may be other agents, goals, and obstacles. The neighborhood information is collected into a graph observation vector $x_j$ that is then passed into the GNN. Graph message passing can provide an encoded representation of goals outside the sensing distance.
For each agent $i$, $x^{(i)}_j=[p^j_i, v^j_i, p^{\mathrm{goal}_1,j}_i, \eta^{\textrm{goal}_1}_i, \texttt{entity\_type(j)}]$ where $p^j_i, v^j_i, p^{\mathrm{goal}_1,j}_i$ are the \emph{relative} position, velocity, and position of the closest goal of the entity at node $j$ with respect to agent $i$, respectively. The variable \texttt{entity\_type(j)} $\in \{ \texttt{agent}, \texttt{obstacle}, \texttt{goal}\}$ determines the type of entity at node $j$. The ego observation is combined with the GNN-encoded observation $x^{(i)}_{agg}$ and input to the agent to produce an action.
This communication mechanism is central to the cooperative nature of the task and further strengthens the system's ability to scale and adapt to various conditions.

\subsubsection{Training reward}
\label{sssec:training_reward}

At every timestep, each agent gets the distance-based reward to an assigned goal, ${\mathcal{R}}_\mathrm{dist}(s_t, a^{(i)}_t)$. When an individual agent reaches their assigned goal (indicated by $\rho$), it receives a one-time goal-reaching reward $\mathcal{R}_\mathrm{goal}(s_t, a^{(i)}_t)$. $\rho$ is 1 if the agent reached the assigned goal and was previously not at the assigned goal; otherwise, 0. In the models that are tuned with fairness metric, we add the fairness reward $\mathcal{R}_\mathrm{fair}(s_t, a^{(i)}_t)$. We also penalize agents colliding with other agents or obstacles in the environment using a collision penalty $-C$. $\kappa$ is a 0/1 variable that indicates if an agent collided with another agent or an obstacle.
The total reward at every timestep then becomes 
\begin{equation}
    \mathcal{R}_\mathrm{total}(s_t, a^{(i)}_t) = {\mathcal{R}}_\mathrm{dist}(s_t, a^{(i)}_t) + \mathcal{R}_\mathrm{fair}(s_t, a^{(i)}_t) + \rho \mathcal{R}_\mathrm{goal}(s_t, a^{(i)}_t) - \kappa C
\label{eq:totalreward}
\end{equation}

\subsubsection{Model variants}
\label{ssss:models}
We use all goal assignment schemes and add the fairness reward to create four models for the coverage navigation scenario, varying using the optimal or fair assignment and including and not including the fairness reward in Eqn. \ref{eq:fair_reward}.
We also train a model that has goals assigned randomly to the agent to investigate the effect of random goal assignments on fairness. 
The four models trained, along with their per-step reward structures, are,

\begin{compactenum}
    \item Random goal assignment with no fairness reward (RA). This model serves as our baseline. 
        \begin{itemize}
            \item$\mathcal{R}_\mathrm{total} = {\mathcal{R}}^{\mathrm{RA}}_\mathrm{dist} + \rho \mathcal{R}_\mathrm{goal} - \kappa C$
        \end{itemize}
    \item Optimal distance cost goal assignment with no fairness reward (OA)
        \begin{itemize}
            \item$\mathcal{R}_\mathrm{total} = {\mathcal{R}}^{\mathrm{OA}}_\mathrm{dist} + \rho \mathcal{R}_\mathrm{goal} - \kappa C$
        \end{itemize}
    \item Fair goal assignment with no fairness reward (FA)
        \begin{itemize}
            \item $\mathcal{R}_\mathrm{total} = {\mathcal{R}}^{\mathrm{FA}}_\mathrm{dist} + \rho \mathcal{R}_\mathrm{goal} - \kappa C$
        \end{itemize}

    \item Fair goal assignment with fairness reward (FA+FR)
        \begin{itemize}
            \item$\mathcal{R}_\mathrm{total} = {\mathcal{R}}^{\mathrm{FA}}_\mathrm{dist} + \mathcal{R}_\mathrm{fair} + \rho \mathcal{R}_\mathrm{goal} - \kappa C$
        \end{itemize}
\end{compactenum}
We do not consider the model in which agents are not assigned specific goals during training, as this results in agents clustering around particular goals and consequently leads to low success rates.

We train models using three agents, three goals, and three obstacles in a fixed environment size. At every time step, the appropriate assignment function determines a goal for each agent based on the current position of all agents and the location of all goals in the environment. This assigned goal is used to obtain the distance-based reward $\mathcal{R}_\mathrm{dist}(s_t, a^{(i)}_t)$. Additionally, for models training using the fairness reward, i.e.,  (FA+FR), the fairness metric is calculated at every time step by considering the distance traveled by all agents from the start of the episode. This is used to provide the fairness reward $\mathcal{R}_\mathrm{fair}(s_t, a^{(i)}_t)$ to the agents. Agents utilize the goal rewards and the fairness reward to learn fair behavior. 

\subsection{Decentralized execution framework}
\label{ssec:execution_framework}

During the evaluation, at the start of each episode, we initialize $N$ agents and $N$ goals randomly in the environment. Each agent can go to any goal in the environment.  Similar to the training setup, the agents have their local observation vector and the neighborhood graph network (as described in Sec. \ref{ssec:observations}).  The policy provides an action for each agent at every time step. 

However, unlike in training, we do not assign any goals; rather, the agents rely on their local observations and the goal assignments learned during training to navigate to the available goals. We also do not provide any distance reward, goal-reaching reward, or fairness reward, as agents do not rely on a centralized critic during execution. 
When an agent reaches a goal, we mark it as `done' until the end of the episode. The episode ends when either all agents are done or the last time step for an episode is reached.

\section{Experiments}
\label{sec:results}
This section presents the main findings of the approach discussed in Section \ref{sec:methods}. We begin with a description of our specific navigation problem setting and introduce our evaluation metrics. We extend our experiments to formation scenarios and investigate how congestion and crowding affect these metrics. All codes and model weights used to evaluate the results are in the open-sourced code-base\footnote{\href{https://github.com/Jaroan/Fair-MARL}{Code base: https://github.com/Jaroan/Fair-MARL}} and the associated Readme.

\subsection{Problem settings}
\label{ssec:prob_setting}

We evaluate models trained with a fair assignment to the optimal distance cost assigned and randomly assigned models on coverage tasks in which agents can navigate to any goal. We want to assess the tradeoff between agents navigating in a fair manner and maintaining efficiency by minimizing the total travel distance. We vary the number of agents and total entities in the environment in our experiments.

\begin{figure}[ht]
    \centering  \hfill
    \begin{subfigure}[b]{0.245\textwidth}
       \includegraphics[width=\textwidth]{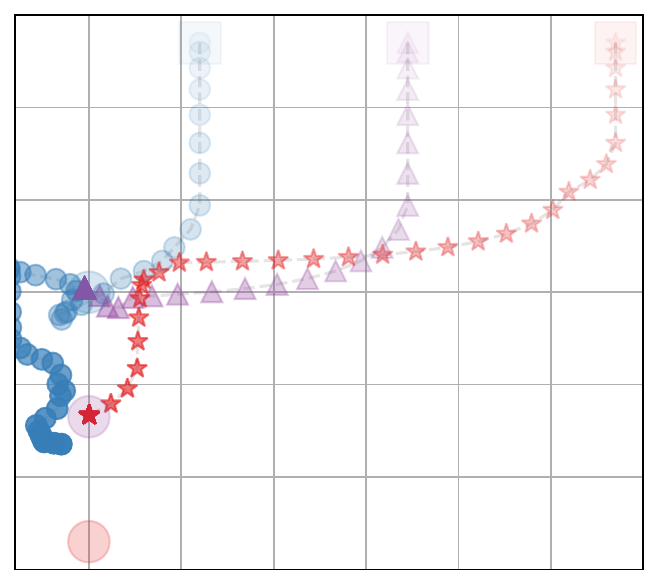}
       \caption{RandomAssign (RA)}
        \label{fig:randomAssignNoFairRew}
    \end{subfigure}\hfill
    \begin{subfigure}[b]{0.23\textwidth}
        \includegraphics[width=\textwidth]{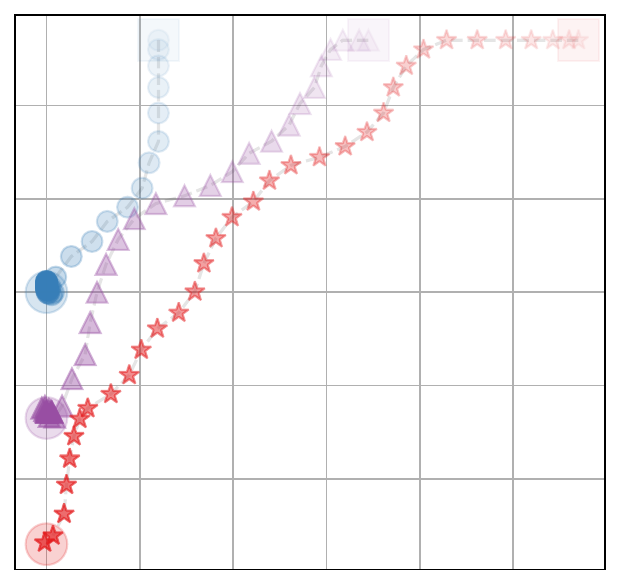}
         \caption{OptAssign (OA)}
        \label{fig:baseline_traj}
    \end{subfigure} \hfill        
        \begin{subfigure}[b]{0.233\textwidth}
        \includegraphics[width=\textwidth]{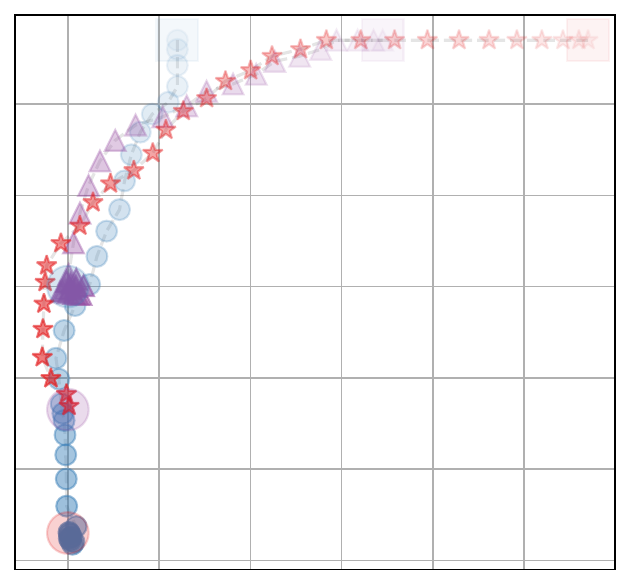}
            \caption{FairAssign (FA)}
        \label{fig:FairAssignNoFairRew}
    \end{subfigure}\hfill
\begin{subfigure}[b]{0.233\textwidth}
\includegraphics[width=\textwidth]{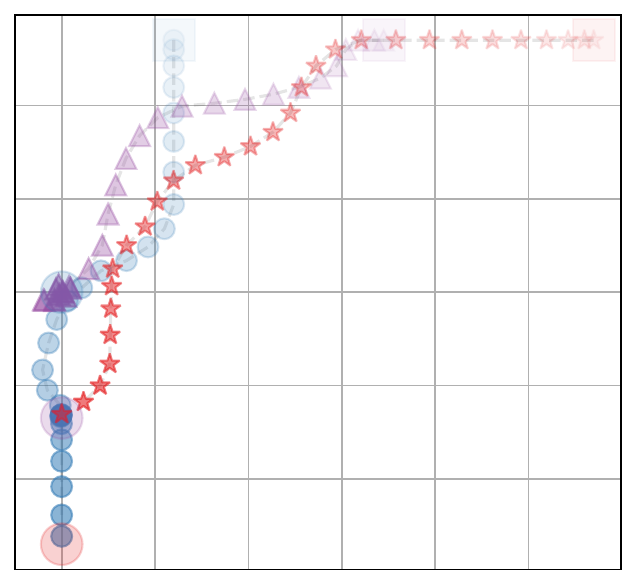}
       \caption{FairAssign, FairRew (FA+FR)}
    \label{fig:FairAssignFairRew}
    \end{subfigure}\hfill
    \includegraphics[width=.5\textwidth]{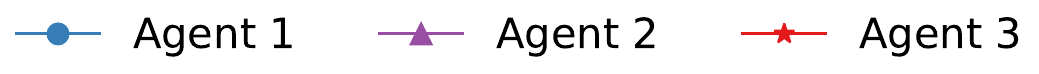}
    \caption{Visualization of behaviors of the four navigation models with and without fairly assigned goals and fairness metric rewards. The agents start from the upper half of the environment and navigate to goals located on the bottom left part of the environment. The darker shades indicate newer states in the trajectories traveled by each agent, and the lighter circles indicate earlier states.
}
    \label{fig:nav_trajs}
\end{figure}

An example evaluation scenario of these models for an episode is shown in Fig. \ref{fig:nav_trajs}. Three agents are initialized in the upper portion of the environment, and the three goals are located in the lower left corner. Agents navigate to the goals based on the different policies they are trained on, and the colored dots show their trajectories. An optimal distance assignment is one where the leftmost agent (Agent 1) selects its nearest goal, the middle agent (Agent 2) chooses the goal in the middle, and the rightmost agent (Agent 3) selects the lower goal. However, this assignment is inherently unfair, as it requires the rightmost agent to cover a significantly longer distance compared to the leftmost agent. A fair assignment would allow the agent located the farthest away to pick its closest goal, i.e., Agent 3 selects the upper goal, and so on.

\begin{table}[ht]
\centering
\caption{Evaluation metrics calculated for the four navigation scenarios shown in Fig. \ref{fig:nav_trajs}. We see that the model that is trained with fair goal assignments and fair rewards (FA+FR) has the best balance of fairness and efficiency (distance traveled).}
\begin{tabular}[b]{|l||c|c|c|c|} \hline 
  Model & Fairness, $\mathcal{F}$ ($\uparrow$ better)  &  Success, $S$\% ($\uparrow$ better) & Episode fraction, $T$ ($\downarrow$ better) & Distance, $D$ ($\downarrow$ better)  \\ \hline \hline 
 RA & \textbf{6.50} & 66.7 & 1.00 & 10.81 \\ \cline{1-5}
 OA & 3.45  & 100.0 & 0.68 & \textbf{9.36}\\ \cline{1-5}

FA & 4.81  & 100.0 & 0.64 & 9.96 \\ \cline{1-5}

 FA+FR & 6.14  &  100.0 & \textbf{0.62}  & 9.82\\  \hline

\end{tabular}

\label{tab:fairness_example}
\end{table}

\subsection{Evaluation metrics}
\label{ssec:eval}

We calculate the following metrics to determine the performance of our method: 
\begin{compactenum}
    \item Fairness, which is the total fairness value (Eqn. \ref{eqn:fairness}) obtained at the end of the episode, denoted by $\mathcal{F}$ (higher the better).
    \item The fraction of an episode time all agents take to reach their goal, denoted $T$ (lower is better). $T$ is set to 1 if any agent does not reach its goal. 
    \item The total distance traveled by the group of agents per episode $D$ (lower is better).
    \item  Success rate as the percentage of agents able to get to unique goals and become `done' denoted by $S$\% (higher is better).

\end{compactenum}
Table \ref{tab:fairness_example} shows the fairness metric $\mathcal{F}$, success rates $S$\%, fraction of episode time $T$, and total distance traveled $D$ by the four navigation models shown in Fig. \ref{fig:nav_trajs}. In the optimal and fair assigned models, the agents successfully reach their goals, setting $S$\% to 100. We see that the model trained with random goal assignments (RA) has the highest fairness metric at the expense of decreased efficiency. The agents also fail to reach all the goals in the environment. However, the model trained with fair goal assignments and a fairness reward (FA+FR) achieves a high fairness metric with a lower total distance traveled. This implies the agents were able to reach the goals using the learned fair assignment along with minimal variance in the individual distance traveled. The model trained with optimal distance cost goal assignments (OA) achieves the most efficient behavior with the lowest distance traveled. But it also has the lowest fairness metric. This shows that (FA and FR) can achieve fair behavior with only a small decrease in overall efficiency.

\subsection{Effect of goal assignments and fairness in reward}

To evaluate the impact of fairness reward and fair goal assignment, we compare the test performance of our four models trained on 3 agents. We run evaluations with 3, 5, 7, and 10 agents over 100 episodes. Our objective was to observe how the fairness reward and goal assignment strategies influence the agents' behavior and performance metrics during execution, where no goals are assigned to a varying number of agents, which is different from what the model was trained on.
\begin{figure}[htbp]
    \centering
    \begin{subfigure}[b]{0.70\textwidth}
        \centering

        \includegraphics[width=\textwidth,trim={0 6 0 0},clip]{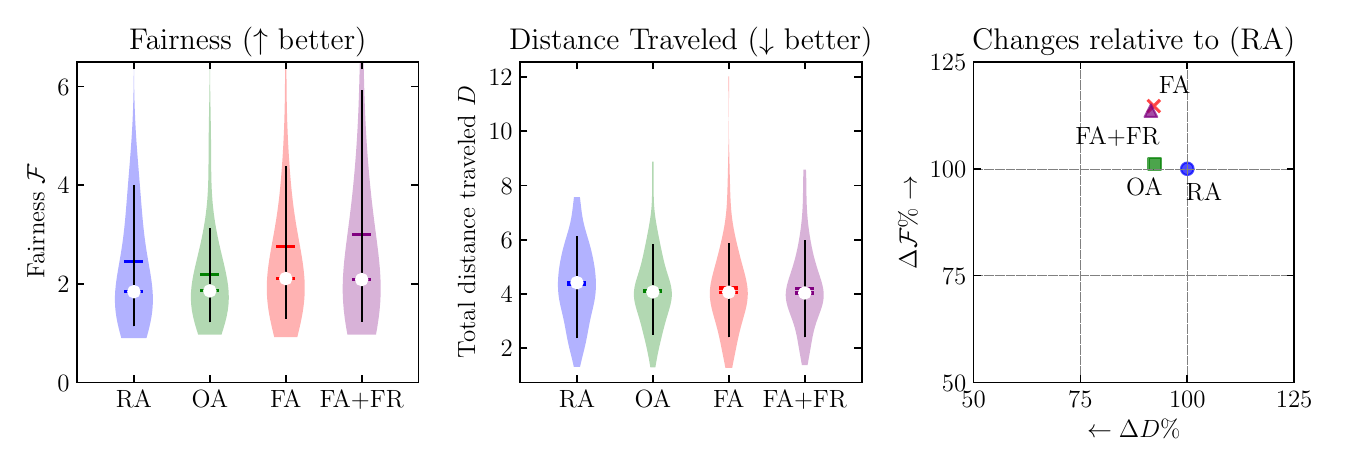}

        \caption{With 3 agents}
        \label{fig:nav3}
    \end{subfigure}\\
    \vspace{13pt}
    \begin{subfigure}[b]{0.70\textwidth}
        \centering
        \includegraphics[width=\textwidth,trim={0 6 0 0},clip]{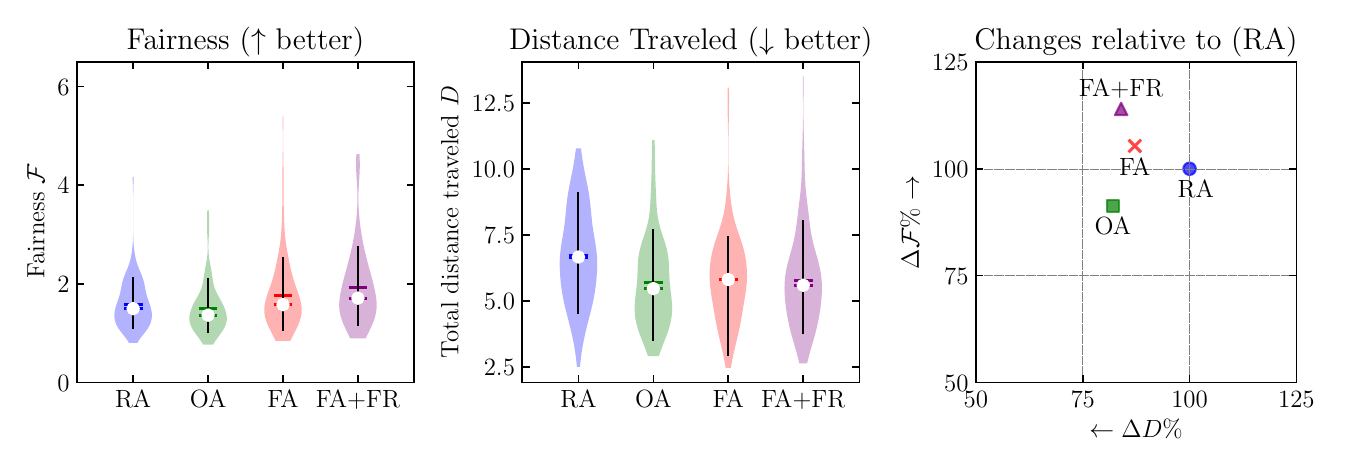}

        \caption{With 5 agents}
        \label{fig:nav5}
    \end{subfigure}
     \\     \vspace{13pt}
    \begin{subfigure}[b]{0.70\textwidth}
        \centering
        \includegraphics[width=\textwidth,trim={0 6 0 0},clip]{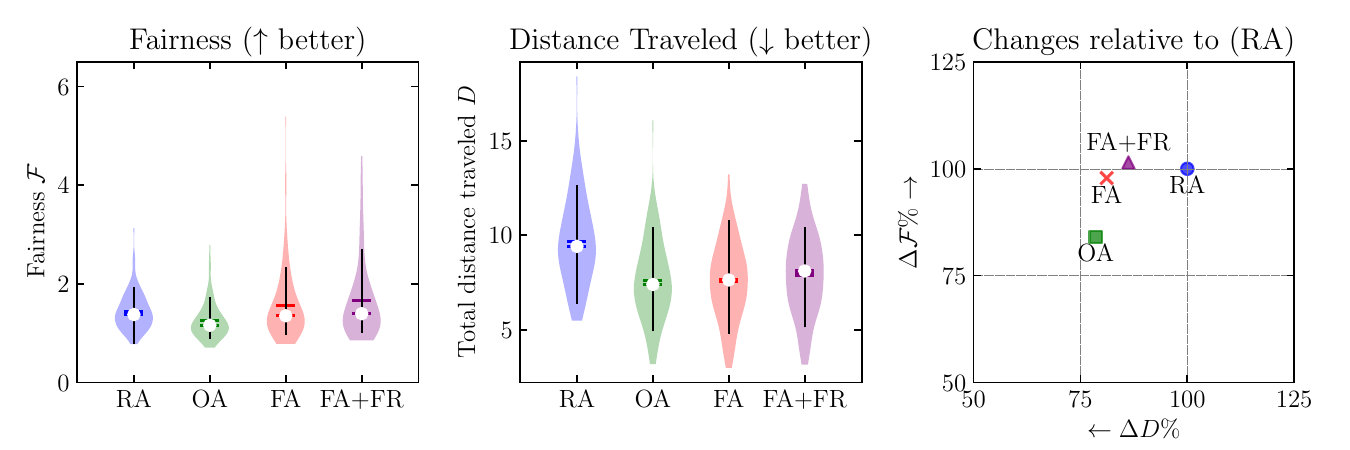}

        \caption{With 7 agents}
        \label{fig:nav7}
    \end{subfigure}\\    \vspace{13pt}
    \begin{subfigure}[b]{0.70\textwidth}
        \centering
        \includegraphics[width=\textwidth,trim={0 6 0 0},clip]{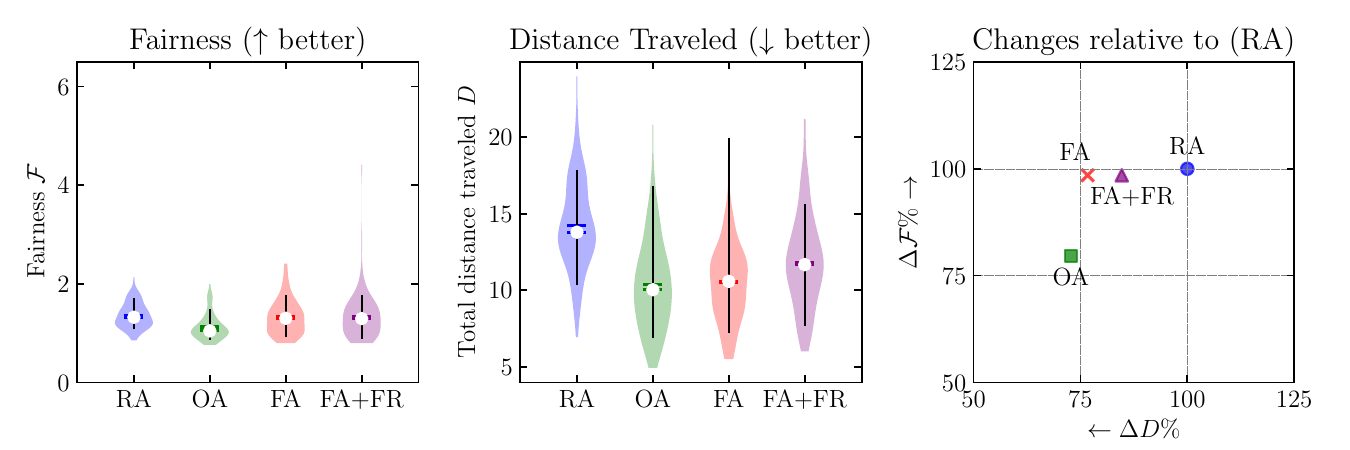}

        \caption{With 10 agents}
        \label{fig:nav10}
    \end{subfigure}
    \caption{The violin plots show the distribution of fairness ($\mathcal{F}$) and the total distance traveled by all agents ($D$) over 100 test episodes for four trained models variants discussed in Section \ref{ssss:models}: 1) Random goal assignments (RA); 2) Optimal distance cost goal assignments (OA); 3) Fair goal assignments  (FA); and 4) Fair goal assignments and a fairness reward (FA+FR). A white circle and tick denote the medians, a plain tick represents the means, and the vertical black lines indicate the 90-10 percentile range. We also show the tradeoffs between fairness and efficiency exhibited by the different models in the rightmost subplot.}
    \label{fig:navigation_distribution}
\end{figure}

Figure \ref{fig:navigation_distribution} illustrates the distribution of $\mathcal{F}$, and $D$ along with the medians, means, and 90-10 percentile ranges for four models in the coverage navigation task: (RA),  (OA), (FA), and (FA+FR) described in Sec. \ref{ssss:models}. We selected these cases to highlight the impact of fair goal assignment and the fairness reward over an efficient and random goal assignment. Table \ref{result_nav_table} shows the median values of $\mathcal{F}$,  $T$, and $D$ and the mean of $S\%$ for all these models. 

\subsubsection{Effect of goal assignments}
The key takeaways regarding the effect of goal assignments are as follows: 
\begin{itemize}
    \item The models that were trained without any goal assignments had very low success rates, i.e., the agents did not cover all the goals. We do not consider these models any further.
    \item Our baseline, the random assignment model (RA), improves the fairness metric but achieves this at the expense of the total distance traveled. 
     Additionally, the model has a lower success rate, indicating that the agents could not learn to cover all the goals effectively within the given episode time compared to the other models.
    \item The model trained with the optimal distance cost goal assignments (OA) achieves the best efficiency, traveling shorter distances. However, this is accomplished by a low fairness metric.
    \item Models trained with min-max fair goal assignment (i.e., (FA) and (FA+FR)) learn to effectively navigate the tradeoffs between efficiency and fairness. This can be attributed to the approach of the min-max fair assignment, which incorporates a distance cost metric, achieving a balance between fairness and efficiency. The (FA, *) models also have higher success rates $S$\% and lower fractions of episode time $T$. These results highlight the efficacy of the min-max fair goal assignment method in optimizing both fairness and efficiency.
    \item Models trained on fair assignments achieve greater fairness than the baseline random assignments models for moderate levels of congestion. We will see later in Section \ref{ssec:congestion} that (RA) is fairer in congested settings.
\end{itemize}

\begin{table*}[htpb]
\caption{Comparison of the four model variants trained with 3 agents and evaluated with m=\{3,5,7,10\} agents for 100 episodes.}
\centering 
\begin{tabular}[b]{|c||c||c|c|c|c|} \hline 
\# agents ($m$)   & Model & $\mathcal{F}$ ($\uparrow$ better)  &  $S$\% ($\uparrow$ better) & $T$ ($\downarrow$ better) & $D$ ($\downarrow$ better)  \\ \hline \hline 
  \multirow{ 4}{*}{3} & RA & 1.84  & 87.7 & 0.68 & 4.42 \\ \cline{2-6}
& OA & 1.86  & 100.0 & 0.40 & 4.08 \\ 
\cline{2-6}
& FA & \textbf{2.11}  & 100.0 & 0.40 & 4.06 \\ \cline{2-6}

& FA+FR & 2.09  &  99.7 & \textbf{0.39}  & \textbf{4.04} \\\hline\hline

\multirow{ 4}{*}{5}
& RA & 1.50  & 86.0 & 0.85 & 6.67 \\ \cline{2-6}

& OA & 1.36 & 99.6 & 0.47 & \textbf{5.47}  \\   \cline{2-6}
& FA & 1.58 & 99.8 & 0.46 & 5.81 \\ \cline{2-6}
& FA+FR  & \textbf{1.71} & 99.8 &  \textbf{0.43} & 5.60 \\
\hline\hline

\multirow{ 4}{*}{7} & RA & 1.38  & 82.6 & 0.94 & 9.42 \\ \cline{2-6}

& OA & 1.16 & 98.3 & 0.59 & \textbf{7.40}  \\  \cline{2-6}
& FA & 1.35 & 98.3 & 0.56 & 7.64  \\ \cline{2-6}
& FA+FR  & \textbf{1.40} & 99.1& \textbf{0.53} & 8.12  \\
\hline\hline

\multirow{ 4}{*}{10} & RA & \textbf{1.32}  & 79.3 & 0.99 & 13.79 \\ \cline{2-6}
& OA & 1.05 & 94.0 & 0.77 & \textbf{10.04}  \\ 
\cline{2-6}
& FA & 1.30 & 96.6 & \textbf{0.73} & 10.58  \\ \cline{2-6}

& FA+FR  & 1.30 & 95.6 &  0.76 & 11.68  \\
 \hline

\end{tabular}
\label{result_nav_table}
\end{table*}

\subsubsection{Effect of fairness reward}
We attempt to increase the fairness of the trained model by including a fairness reward in addition to the min-max fair goal assignment (FA+FR).
While training this model, the fairness metric is checked at every step of the episode. The agents start navigating to the goals, and the fairness reward is applied to the agents in addition to the distance reward. Thus, agents are incentivized to maintain a high fairness metric throughout the episode. From the results, our method (FA+FR) is among the highest fairness metrics providing models, demonstrating the impact of combining a fair assignment with a fairness reward.

We compute the improvement in the median fairness metric $\mathcal{F}$ of our (FA+FR) model to the (OA) model for the 3, 5, 7, and 10 agent scenarios as detailed in Table \ref{result_nav_table}. We also calculate the change in the total distance traveled $D$ for the (FA+FR) model to the (OA) model.
From all the experiments, we find that the fairness metric for the (FA+FR) model has an average of 20.65$\pm$5\% improvement when compared to the (OA) model, with only an average of 6.86$\pm$6\% increase in the total distance traveled by the agents. This shows that training with a fair goal assignment and fairness reward improves overall fairness in the navigation scenario without an extensive tradeoff in efficiency. 

Similarly, we calculate the change in the median fairness metric, $\mathcal{F}$, of our (FA+FR) model compared to the (RA) model across various agent scenarios (3, 5, 7, and 10 agents), using the values in Table \ref{result_nav_table}. We also analyze the corresponding change in the total distance traveled, $D$, for the same models. The results indicate an average of 13.85$\pm$3\% improvement in the total distance traveled by the agent and an average of 5.04$\pm$7\% increase in the fairness metric. The results show that our model (FA+FR) has a notable gain in efficiency over the random baseline, suggesting the benefit of including a structured assignment of goals for agents. The fairness metric also has a modest improvement. The higher standard deviation here indicates the improvement in our model's fairness metric depends on the agent configuration in the environment. When the number of agents is large, the fairness metric values are similar for the (FA+FR) and (RA) models, as seen in Fig. \ref{fig:nav10}.

We observe a different trend as we increase the number of agents in each scenario. Although the (FA+FR) model maintains a high fairness metric for the 7 and 10 agent experiments, the distance traveled is higher than the models trained with just fair or optimal assignments. When the number of agents and goals in an environment increases, an agent may not be able to navigate to its goal as quickly as compared to the 3 agent experiment. The agents experience congestion with a greater agent count in the same environment size. This leads to more collisions and deviations from the shortest paths. This resulted in a decrease in overall fairness and success rates. We further describe this phenomenon in Section \ref{ssec:congestion}.

We compare our model (FA+FR) with the Fair-Efficient Network (FEN) \cite{jiang2019learning} and calculate the median values of $\mathcal{F}$,  $T$, and $D$ and the mean of $S\%$ evaluated for 100 episodes. The values for 3, 5, 7, and 10 agent scenarios are listed in Table \ref{result_FEN_comparison}. (FA+FR) outperforms FEN significantly across all agent counts. FEN also has low success rates with a similar environment setup. This resulted in shorter travel distances, and the time required spanned the entire episode.

\begin{table*}[htpb]
\caption{Comparison of the (FA+FR) model variant with the FEN model \cite{jiang2019learning} trained with 3 agents and evaluated with m=\{3,5,7,10\} agents for 100 episodes.  We note that although the FEN model appears more efficient in terms of the distance metric, the success rates are very low compared to the (FA+FR) mode.}
\centering 
\begin{tabular}[b]{|c||c||c|c|c|c|} \hline 
\# agents ($m$)   & Model & $\mathcal{F}$ ($\uparrow$ better)  &  $S$\% ($\uparrow$ better) & $T$ ($\downarrow$ better) & $D$ ($\downarrow$ better) \\ \hline \hline 
  \multirow{ 2}{*}{3} & FEN & 1.01  & 10.0 & 1.0 & 1.52 \\ \cline{2-6}

& FA+FR & 2.09  &  99.7 & 0.39  & 4.04 \\\hline\hline

\multirow{2}{*}{5}
& FEN & 1.08  & 10.0 & 1.0 & 3.70 \\ \cline{2-6}
& FA+FR  & 1.71 & 99.8 &  0.43 & 5.60 \\
\hline\hline

\multirow{ 2}{*}{7} & FEN & 1.05  & 10.0 & 1.0 & 4.77 \\ \cline{2-6}
& FA+FR  & 1.40 & 99.1& 0.53 & 8.12  \\
\hline\hline

\multirow{2}{*}{10} & FEN & 1.07  & 10.0 & 1.0 & 7.30 \\ \cline{2-6}
& FA+FR  & 1.30 & 95.6 &  0.76 & 11.68  \\
 \hline

\end{tabular}
\label{result_FEN_comparison}
\end{table*}

\begin{figure}[htpb]
    \centering
    \includegraphics[width=0.6\textwidth]{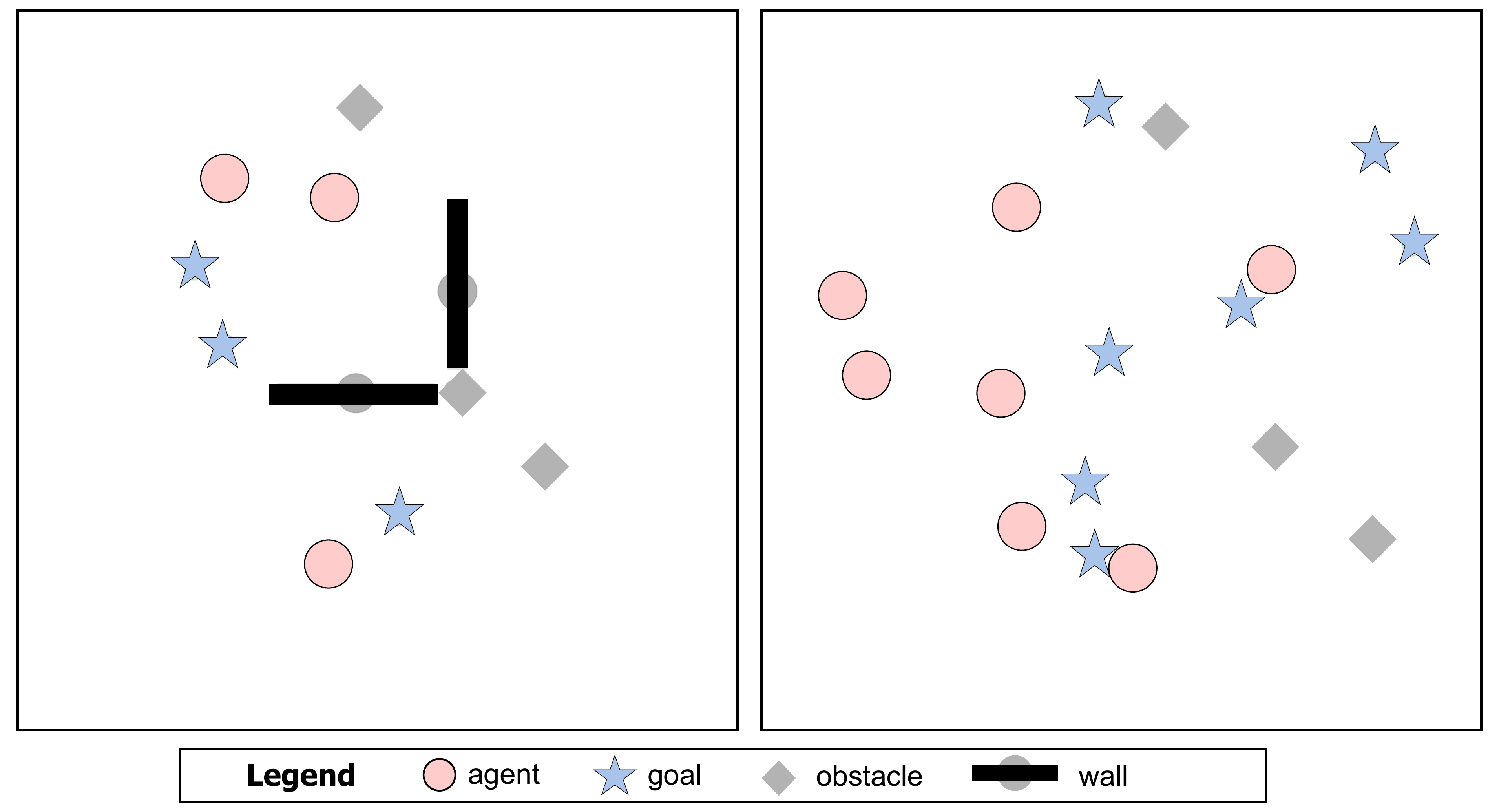}
    \caption{Congestion in the environment: The figure on the left shows an environment with 3 agents along with 3 obstacles and 2 walls. The figure on the right shows the environment with 7 agents and 3 obstacles. The environment is crowded with the increased number of agents, which decreases free space for navigating in straight lines.}
    \label{fig:congestion}
\end{figure}

\subsection{Impact of congestion on overall fairness and efficiency }
\label{ssec:congestion}

In this subsection, we examine the effect of increased agent density on agent performance metrics. We also investigate how obstacles and walls in the environment impact overall fairness. A common challenge in transportation systems is to manage traffic flow in congested areas, such as during urban rush hours. In these scenarios, streets are often packed with vehicles, leading to increased travel times and higher chances of collisions. As the number of agents increases in a fixed environment size, the higher entity density leads to more agent interactions and a greater likelihood of collisions compared to cases with fewer agents. Obstacles and walls can create bottlenecks and restricted pathways, leading to unequal access to resources or destinations for different agents. The left half of Fig. \ref{fig:congestion} shows an environment of three agents with the addition of obstacles and walls. On the right, we see the same environment size with seven agents. The increase in the entity density leads to more agent interactions and increased collisions compared to experiments with fewer agents. 
\begin{figure}[hp]
    \centering
    \begin{subfigure}[b]{\textwidth}
        \centering
        \includegraphics[width=0.9\textwidth,trim={0 5 0 0},clip]{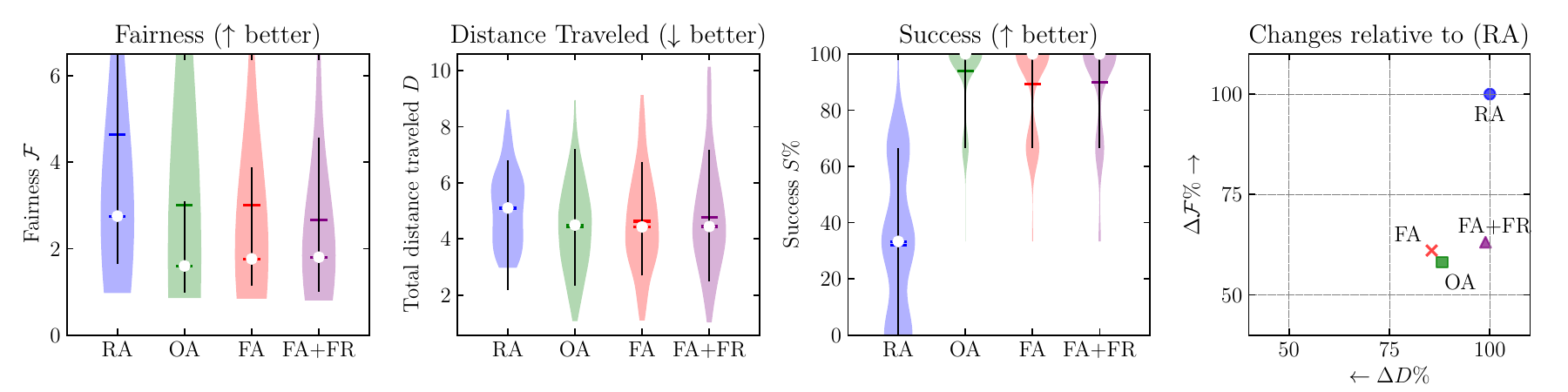}
        \caption{With 3 agents}
        \label{fig:wall3}
    \end{subfigure}\vspace{13pt}
     \\
    \begin{subfigure}[b]{\textwidth}
        \centering
        \includegraphics[width=0.9\textwidth,trim={0 5 0 0},clip]{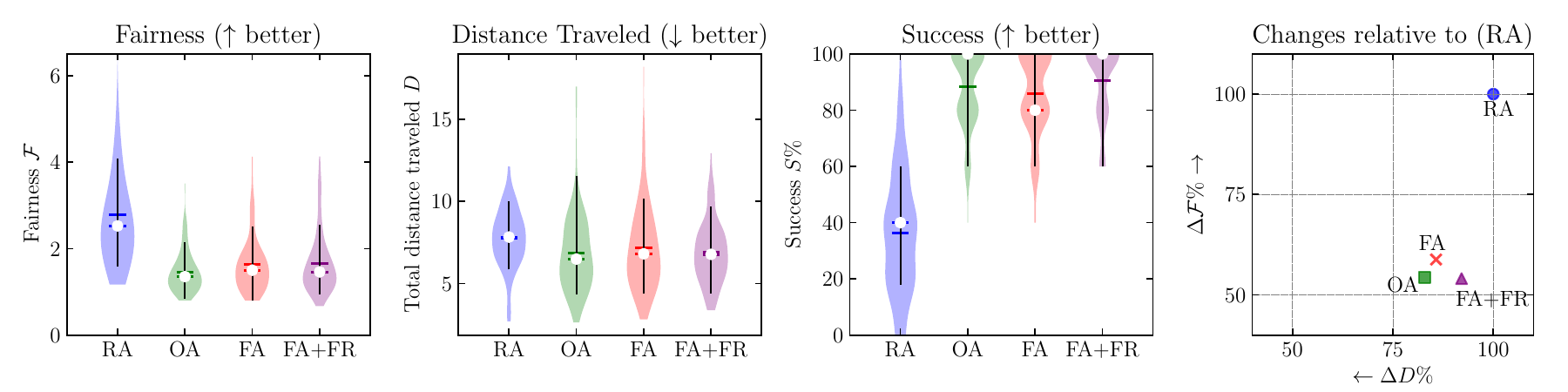}
        \caption{With 5 agents}
        \label{fig:wall5}
    \end{subfigure}\vspace{13pt}
     \\
    \begin{subfigure}[b]{\textwidth}
        \centering
        \includegraphics[width=0.9\textwidth,trim={0 5 0 0},clip]{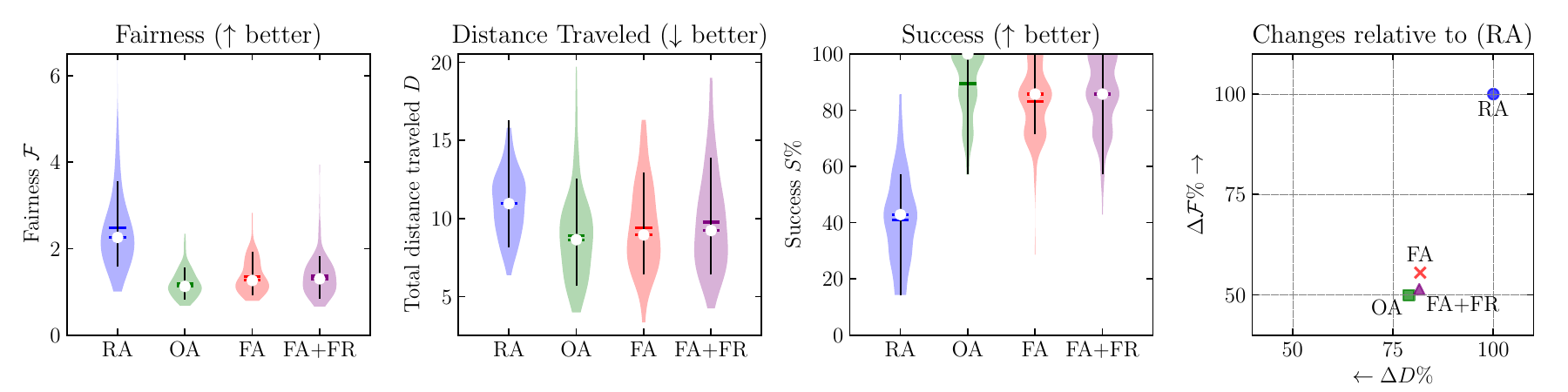}
        \caption{With 7 agents}
        \label{fig:wall7}
    \end{subfigure}
     \\\vspace{13pt}
    \begin{subfigure}[b]{\textwidth}
        \centering
        \includegraphics[width=0.9\textwidth,trim={0 5 0 0},clip]{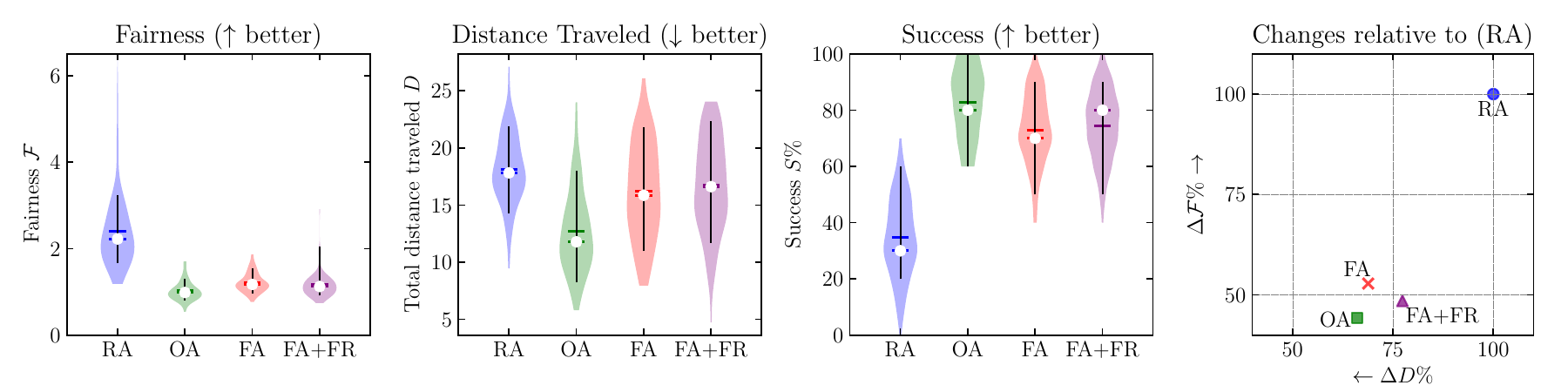}
        \caption{With 10 agents}
        \label{fig:wall10}
    \end{subfigure}
    \caption{Congestion: The violin plots show the distribution of fairness ($\mathcal{F}$), total distance traveled ($D$) and success rates ($S\%$) over 100 test episodes for four trained model variants: 1) Random goal assignments  (RA); 2) Optimal distance cost goal assignments (OA); 3) Fair goal assignments (FA); and 4) Fair goal assignments and a fairness reward (FA+FR).  A white circle and tick denote the medians, a plain tick represents the means, and the vertical black lines indicate the 90-10 percentile range. We also show the tradeoffs between fairness and efficiency exhibited by the different models in the rightmost subplot.}
    \label{fig:navigation_with_wall_obst}
\end{figure}

Figure \ref{fig:navigation_with_wall_obst} shows the distribution of $\mathcal{F}$, $D$, and $S\%$ for different numbers of agents in the presence of obstacles and walls. We see a trend similar to the previous experiments where a higher number of entities create a congested environment. With additional obstacles and walls, agents must frequently adjust their paths, resulting in increased travel distances. As the environmental complexity increases, higher fairness metrics come at the expense of the total distance traveled. The figure shows that the (FA+FR) model maintains a high fairness metric but travels larger distances to achieve this. The (FA) model also performs similarly in fairness and distance traveled but suffers from a lower success rate. (OA) model travels the least total distance. The randomly assigned model surpasses the other three models in terms of the fairness metric; however, this advantage is offset by an increase in total distance and a significant drop in the success rates.

When analyzing the behavior of the agents in this congested scenario more broadly, we see that during the initial time steps, agents observe nearby goals, but these may not correspond to their fairly assigned goals. As agents are farther away from their fair goals, they may prioritize receiving a fairness reward over a goal reward. This could result in agents moving around the goals to reduce the deviation in the distance traveled and optimize their fairness metrics. Specifically, in the (FA+FR) model, agents deliberately loiter to improve the fairness metric to receive a higher fairness reward to compensate for the lack of an easy, earlier goal reward. This behavior highlights the complexity involved in achieving fairness while maintaining efficiency in large environments with many entities.

\subsection{Formation scenario}
\label{ssec:formation}

\begin{figure}[h]
    \centering
        \includegraphics[width=\linewidth]{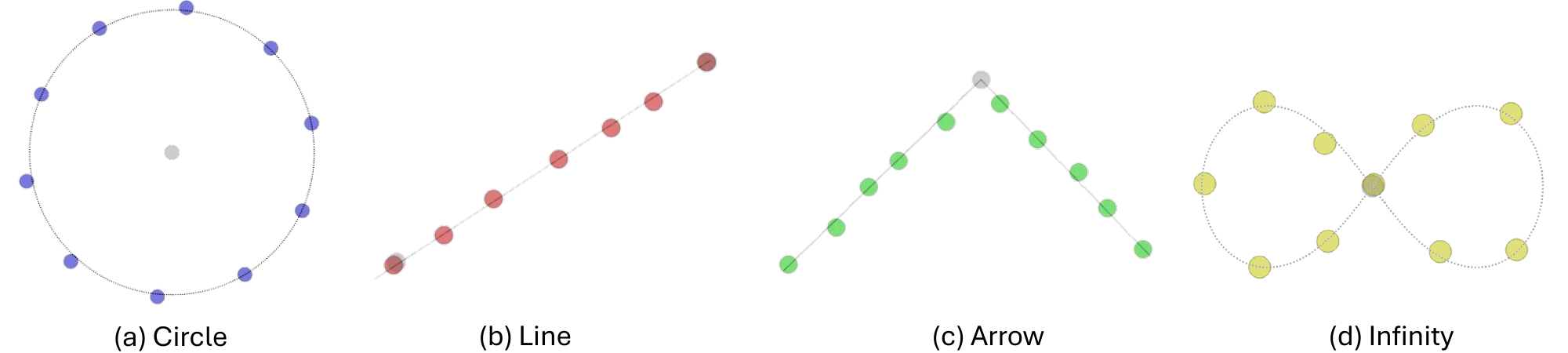}
    \caption{Different formations: Various shapes are created using a set of "expected positions" around one or two landmark positions. The agents arrange themselves on or near these expected positions to form different shapes.}
    \label{fig:form_types}
\end{figure}

Apart from navigation scenarios, our approach can be extended to agents coordinating and forming various shapes. 
We consider four formation shapes: a circle, a line, an arrow, and a lemniscate curve (or, the infinity shape, with parametric equations $ x = a \mathrm{sin}\ t/(1+\mathrm{cos}^2t)$ and $y = a \mathrm{sin}\ t\mathrm{cos}\ t/(1+\mathrm{cos}^2\ t)$, where $a$ is the half-width), as shown in Fig. \ref{fig:form_types}. The shapes have either a single or pair of landmarks, which could be the center or the endpoints. The shapes are discretized into multiple points equal to the number of agents used for the experiment. These points are the "expected positions" the agents can occupy to form the shapes. Similar to the observation vector in the navigation scenario (Sec.\ref{ssec:observations}), the agents are provided the nearest expected position. During training, the reward function assigns positions based on the fair goal assignment scheme or the optimal distance cost goal assignment scheme. These assignments are utilized for agents to learn to reach unique positions and arrange themselves in an equidistant manner. The agents receive a distance-based reward at every time step, and once they get to their assigned position, they receive a goal reward. We also include the fairness reward and train three models of fairness-informed experiments similar to the navigation scenario, namely (FA+FR), (FA), and (OA). We do not consider the randomly assigned goals model (RA) because of its low success rates in the coverage navigation task.

 As the "expected positions" are independent of the shape used in the training, this method can allow agents to come into any desired formation based on the location of the landmark(s). We can utilize a model trained on a circle formation to create any other formation using a different set of discretized expected positions. However, unlike the goal navigation scenario, each expected position is not initialized like a goal landmark entity. In the navigation scenario, agents travel to fixed goal targets. In contrast, in the formation scenario, there are multiple position choices for an agent to travel to, depending on the total number of agents and the degree of symmetry of the formed shape.
 Therefore, the expected positions are only passed into the graph observation vector as a part of the goal position for an $\texttt{agent}$ entity. Only the landmarks present in the environment are processed like the $\texttt{goal}$ entity type, unlike the target navigation case where all goal positions were modeled as $\texttt{goal}$ entities. Finally, we utilize the same observation function and the assignment of goals from the navigation scenario to maintain a consistent model architecture.

 \subsubsection{Evaluation of formation}
 During deployment, we do not assign specific target positions to the agents. Instead, agents rely on the trained goal assignments to arrange themselves on the expected positions observed at each time step.

To evaluate the performance of the formation task, we use the following success metric tailored for each formation.
\begin{compactenum}
    \item Circle: Agents must reach within a threshold distance of a given radius to a central landmark.
    \item Line: Agents must reach within a threshold distance along the line connecting the two landmarks representing the endpoints of the line. 
    \item Arrow: Agents arrange themselves on the two tails of the arrowhead, with the tip represented by the landmark.
    \item  Infinity: The landmark is placed at the midpoint of the formation. Agents arrange themselves on the two lobes, approximated as two circles for ease of calculation. We measure the agent's distance from either of the two centers, which must be within a threshold radius.
\end{compactenum}

\begin{figure}[htbp]
    \centering
    \begin{subfigure}[b]{\textwidth}
        \centering
        \includegraphics[width=0.45\textwidth]{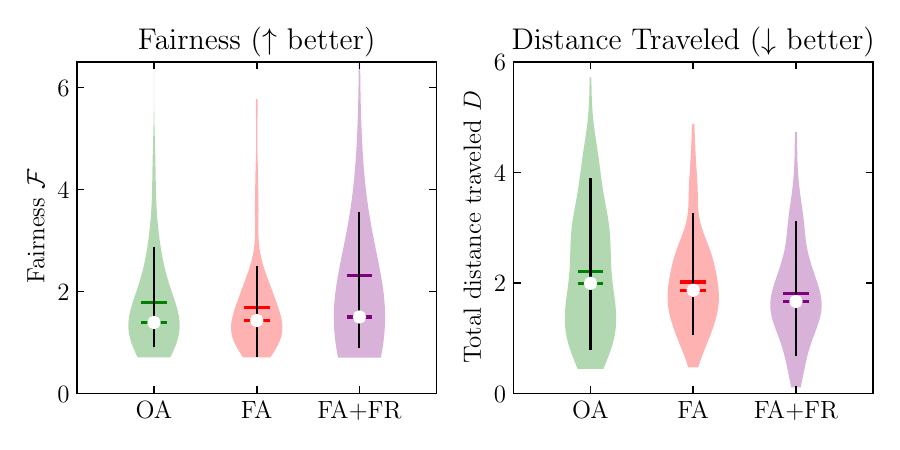}

        \caption{With 3 agents}
        \label{fig:circle3}
    \end{subfigure}
     \\
    \begin{subfigure}[b]{\textwidth}
        \centering
        \includegraphics[width=0.45\textwidth]{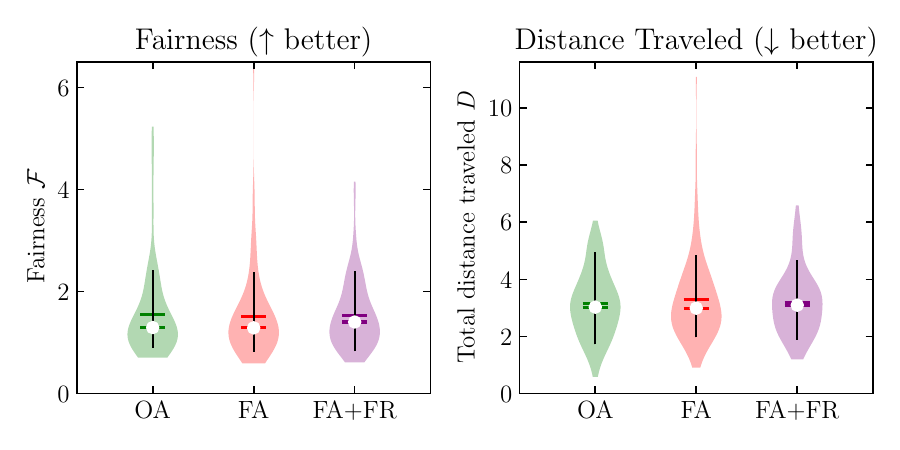}

        \caption{With 5 agents}
        \label{fig:circle5}
    \end{subfigure}
     \\
    \begin{subfigure}[b]{\textwidth}
        \centering
        \includegraphics[width=0.45\textwidth]{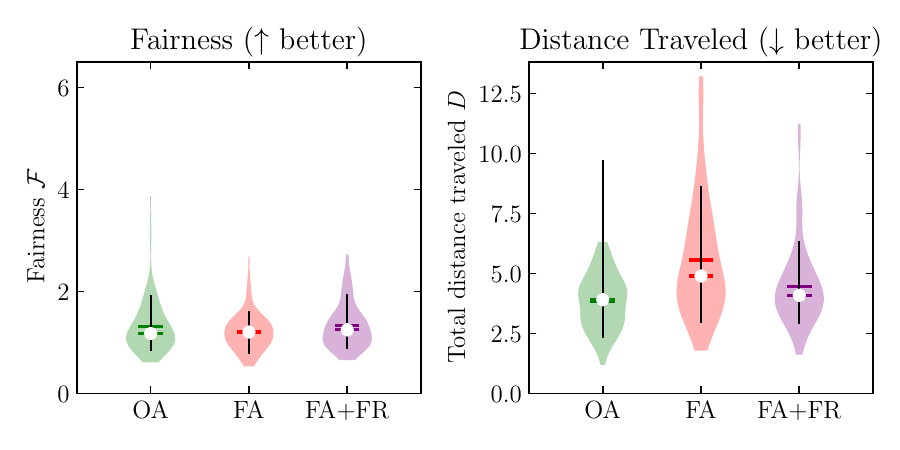}

        \caption{With 7 agents}
        \label{fig:circle7}
    \end{subfigure}
     \\
    \begin{subfigure}[b]{\textwidth}
        \centering
        \includegraphics[width=0.45\textwidth]{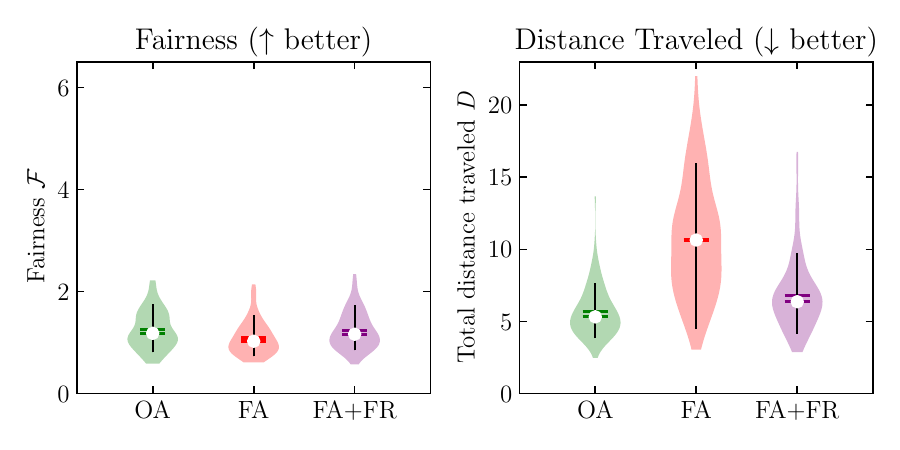}

        \caption{With 10 agents}
        \label{fig:circle10}
    \end{subfigure}
    \caption{Circle Formation: The violin plots show the distribution of fairness ($\mathcal{F}$) and the total distance traveled ($TD$) over 100 test episodes for three trained model variants: 1) Optimal distance cost goal assignments (OA); 2) Fair goal assignments (FA); and 3) Fair goal assignments and a fairness reward (FA+FR). A white circle and tick denote the medians; a plain tick represents the means, and the vertical black lines indicate the 90-10 percentile range.}
    \label{fig:circle_formation}
\end{figure}

Agents are initialized at random positions and tasked to arrange themselves in various formation shapes within the given episode time. We evaluate our four fairness-informed models described in Sec. \ref{ssec:prob_setting} over 100 episodes. Similar to previous evaluations, we do not assign any target positions, and the agents rely on the assignments they have learned during training to create a formation. We also do not provide any distance or goal-reaching rewards as we do not rely on a centralized critic during execution. We note when agents reach their first position and mark them as `done' till the end of an episode.
Figure \ref{fig:circle_formation} shows the distribution of $\mathcal{F}$ and $D$ along with the medians, means, and 90-10 percentile ranges for the three models: (FA+FR), (FA) and (OA). 

Table \ref{result_circle_table} shows the median values of $\mathcal{F}$, $T$, and $D$ and the mean of $S\%$ for all three models. The trends observed in the fairness metric are analogous to those in the navigation scenario. Model (FA+FR) has among the highest fairness metrics, with a higher success metric and low episode fraction time. We also see the total distance traveled is comparable to the optimal assignment cases without any fairness metric used in the reward. This indicates that fairness is achieved without a substantial compromise in efficiency, as the distances required for fair behavior do not increase significantly. From all the circle formation experiments, we find that the (FA+FR) model has an average improvement of 5.55$\pm$3\% in fairness when compared to the (OA) model. However, this comes at the expense of a 2.96$\pm$12\% decrease in efficiency (as measured by the total distance traveled by the agents). 

Table \ref{result_all_formation} presents a comparison of the test performance for models trained on a circle formation with 3 agents and evaluated on the circle, line, arrow, and infinity formation scenarios with 5 agents. The results indicate the models have high success rates on all formations, showing the generalizability of the method. We note the min-max fair assigned models (FA+FR) and (FA) have high fairness metrics. In addition, the model (FA+FR) is able to tradeoff being fair with comparable efficiency as the optimal distance cost model (OA) while maintaining low episode fraction times.

The necessity for a separate formation model arises from the need to address more nuanced control problems. Unlike navigation problems, which primarily involve reaching predefined targets, formation problems require each agent to consider the relative positioning and approach such that they can ensure unique structures as a team. In our method, the agents implicitly learn the relative positioning around the central landmark using the graph network. We increase the difficulty and constraints on movement and positioning, which can lead to more sophisticated and robust multi-agent coordination strategies.

\begin{table*}[ht]
\caption{Comparison of the circle formation model trained with 3 agents and evaluated with m=\{3,5,7,10\} agents for 100 episodes.}
\centering 
\begin{tabular}[b]{|c||c||c|c|c|c|} \hline 
\# agents ($m$)   & Model & $\mathcal{F}$ ($\uparrow$ better)  &  $S$\% ($\uparrow$ better) & $T$ ($\downarrow$ better) & $D$ ($\downarrow$ better)  \\ \hline \hline 
 \multirow{ 4}{*}{3} & OA & 1.39  & 100.0 & 0.28 & 1.99 \\   \cline{2-6}
& FA & 1.43 & 99.7 & 0.25 & 1.87 \\ \cline{2-6}

& FA+FR & \textbf{1.50}   &  100.0 & \textbf{0.22}  & 1.67\\\hline\hline

\multirow{ 4}{*}{5} & OA & 1.29 & 100.0 & 0.31 & 3.03  \\  \cline{2-6}
& FA & 1.29 & 96.8 & 0.32 & \textbf{2.99} \\ \cline{2-6}

& FA+FR  & \textbf{1.40} & 99.8 &  \textbf{0.30} & 3.10 \\ \hline\hline

\multirow{ 4}{*}{7} & OA & 1.18 & 98.8 & \textbf{0.33} & \textbf{3.91}  \\  \cline{2-6}
& FA & 1.20 & 93.0 & 0.48 & 4.90  \\ \cline{2-6}

& FA+FR  & \textbf{1.25} & 99.6& 0.36 & 4.10  \\ \hline\hline

\multirow{ 4}{*}{10} & OA & \textbf{1.18} & 96.8 & \textbf{0.37} & \textbf{5.32}  \\ \cline{2-6}
& FA & 1.03 & 82.8 & 0.73 & 10.64  \\ \cline{2-6}

 & FA+FR  & 1.16 & 98.9 &  0.43 & 6.37  \\ \hline

\end{tabular}
\label{result_circle_table}

\end{table*}
\begin{table*}[ht]
\caption{Comparison of all formation models trained with 3 agents and evaluated with 5 agents for 100 episodes.}
\centering 
\begin{tabular}[b]{|c||c||c|c|c|c|} \hline 
Formation & Model & $\mathcal{F}$ ($\uparrow$ better)  &  $S$\% ($\uparrow$ better) & $T$ ($\downarrow$ better) & $D$ ($\downarrow$ better)  \\ \hline \hline 
\multirow{ 4}{*}{Circle} & OA & 1.29 & 100.0 & 0.31 & 3.03  \\  \cline{2-6}
& FA & 1.29 & 96.8 & 0.32 & \textbf{2.99} \\ \cline{2-6}

& FA+FR  & \textbf{1.40} & 99.8 &  \textbf{0.30} & 3.10 \\ \hline\hline

\multirow{ 4}{*}{Line} & OA & 1.34 & 97.8 & 0.54 & \textbf{6.12}  \\  \cline{2-6}
& FA & \textbf{1.70} & 100.0 & \textbf{0.73} & 13.16 \\ \cline{2-6}

& FA+FR  & 1.42 & 96.2 &  \textbf{0.73} & 8.06 \\ \hline\hline

\multirow{ 4}{*}{Arrow} & OA & 1.29 & 100 & \textbf{0.49} & \textbf{4.96}  \\  \cline{2-6}
& FA & \textbf{1.36} & 86.0 & 0.92 & 10.24  \\ \cline{2-6}

& FA+FR  & \textbf{1.36} & 96.0 & 0.58 & 6.08  \\ \hline\hline

\multirow{ 4}{*}{Infinity} & OA & 1.12 & 99.4 & 0.61 & 6.16  \\ \cline{2-6}
& FA & \textbf{1.31} & 94.4 & 0.68 & 7.26  \\ \cline{2-6}

 & FA+FR  & 1.22 & 99.6 &  \textbf{0.57} & \textbf{6.02}  \\ \hline

\end{tabular}
\label{result_all_formation}
\end{table*}
\subsection{Limitations}
\label{ssec:limitations}
We identify the following limitations of the proposed approach in its current form:
 
\begin{compactenum}
    \item \textbf{Increased fairness at the expense of efficiency:} As the number of agents increases, the difficulty of navigating to easily reachable and observable goals also increases. In such cases, agents often prioritize improving the fairness metric over getting to the goal, which can lead to suboptimal behavior like loitering. Proper tuning and balancing reward proportions ensure that agents are incentivized to reach their assigned goals while maintaining fairness in a scenario-independent manner. 

    \item \textbf{Scalability challenges in dense environments:} Maintaining fairness becomes significantly more challenging as the number of agents grows. Increased agent density and the presence of obstacles introduce difficulties in achieving both fair and efficient outcomes. While fair goal assignment and fairness rewards can improve fairness metrics, they often result in increased travel distances and more complex navigation behaviors. This, in turn, can lead to higher computational costs, hindering real-time applications in dynamic environments.

    \item \textbf{Predefined formations:}
    The current implementation of the formation scenarios assumes that predefined expected positions are set up for each environment. Inter-agent interactions and the effect of the landmark(s) are captured implicitly through the GNN. While the method successfully adapts to different shapes, creating intelligent formations will require greater consideration of inter-agent interactions and landmark locations. 

\end{compactenum}

\section{Conclusions and Future Work}
\label{sec:conclusions}

In this work, we addressed the problem of fair multi-agent navigation using multi-agent reinforcement learning. We utilized the reciprocal of the coefficient of variation of the distances traveled by different agents as a fairness metric. Our proposed model (FA+FR) incorporated training agents using min-max fair distance goal assignments along with a reward term to incentivize fairness during their movement. Our results show agents can learn fair assignments without needing to significantly sacrifice efficiency. Additionally, our model achieved almost perfect goal coverage even when tested on a larger number of agents than training, showing the scalability of our approach. We compared the (FA+FR) model to ones trained by assigning goals using an optimal distance cost (i.e., which optimized efficiency) and a baseline model that used randomized goal assignments. 

For goal coverage scenarios, in comparison to the baseline model trained with random assignments, our model resulted in an average of 14\% improvement in efficiency (as measured by the total distance traveled by the agents) and a 5\% average increase in fairness.
Additionally, we obtained an average of 21\% improvement in fairness with only an average of 7\% decrease in the efficiency with the (FA+FR) model as compared to a model trained on optimal assignments and no fairness in the rewards. However, as the environment grew more congested, agents sometimes traveled longer distances simply in order to improve the fairness metric, resulting in an unnecessarily large decrease in efficiency.

Our approach enables a greater level of decentralization with less dependence on a centralized oracle. As an added benefit, we obtain some privacy by not requiring all agents and goal positions to be known before the navigation starts. We also implemented ``death masking'' to prevent collision forces from influencing the goal-reaching behaviors.
We demonstrated the extensibility of our method by applying it to various formation tasks. We showed that our method is able to generalize to different formation shapes and achieve complete coverage without retraining the models.
We found the overall fairness metric becomes lower as we increase the number of agents.

Interesting directions for further investigation include the consideration of other measures of fairness, the development of heuristics for large-scale, highly-congested environments, and extending the proposed approach to the fair and efficient creation of dynamic formations using multiple vehicles. The latter could, for example, be achieved by designating specific agents as leaders and others as followers, thereby reducing reliance on static external landmarks. Doing so would enhance adaptability to dynamic environments and enable more autonomous formation control.
Practical deployment introduces additional complexities such as sensor noise, communication constraints, hardware limitations, and dynamic environmental conditions. Future work would include considering robustness to real-world noise and uncertainties caused by sensor errors and delayed communication. Enhanced decentralized cooperation would allow agents to take action intelligently during periods of little to no communication due to distance or other obstacles. Expanding to simulators that incorporate physical constraints of real-world hardware deployment will allow us to test the limitations of our approach, train in diverse scenarios, and adapt to unforeseen environments or dynamically changing conditions.



\begin{acks}
This work was supported in part by NASA under grant \#80NSSC23M0220 and the University Leadership Initiative (grants \#80NSSC21M0071 and \#80NSSC20M0163), but this article solely reflects the opinions and conclusions of its authors and not any NASA entity. J. Aloor was also supported in part by a Mathworks Fellowship. The authors would like to thank the MIT SuperCloud \citep{supercloud} and the Lincoln Laboratory Supercomputing Center for providing high performance computing resources that have contributed to the research results reported within this paper.
\end{acks}

\bibliographystyle{ACM-Reference-Format}
\bibliography{references}









\end{document}